\documentclass[amsmath,amssymb]{article}
\def\@preprint{}
\newcommand\preprint[1]{\@preprint{\hfill #1}}

\newcommand{\maketitleNew}[1]{

	\setcounter{page}{0}
	\noindent{\small\scshape}\@preprint\par
	\preprint{\begin{tabular}{r}MITP/20-062\\ LTH 1247 \end{tabular} }

	{\let\newpage\relax\maketitle}
	\maketitle
	\thispagestyle{empty}
}
\pdfoutput=1
\usepackage{authblk}
\usepackage{abstract}
\usepackage[footnotesize]{caption}
\usepackage{subcaption}
\usepackage{cite}
\usepackage{amsmath}
\usepackage{amssymb}
\usepackage{graphicx} 
\usepackage[latin1]{inputenc}
\usepackage{mathrsfs}
\usepackage[dvipsnames]{xcolor}
\usepackage{hyperref}
\usepackage[hmarginratio=1:1,top=32mm,columnsep=20pt]{geometry}
\usepackage{paralist}
\usepackage{multirow}
\usepackage{multicol}
\usepackage{blindtext}
\usepackage{upgreek}
\usepackage{soul}

\newcommand{\bea}{\begin{eqnarray}}
\newcommand{\eea}{\end{eqnarray}}
\newcommand{\be}{\begin{equation}}
\newcommand{\ee}{\end{equation}}
\newcommand{\ba}{\begin{array}}
\newcommand{\ea}{\end{array}}

\def\gsim{\mathrel{\rlap{\lower4pt\hbox{\hskip1pt$\sim$}}
    \raise1pt\hbox{$>$}}}

\renewcommand\Affilfont{\normalsize\itshape}

\title{\fontsize{14pt}{10pt}\selectfont 
	\textbf{Testing CP Properties of Extra Higgs States \\ at the HL-LHC}
	}
\author[1]{\normalsize Stefan~Antusch\thanks{\texttt{stefan.antusch@unibas.ch}}}
\author[2]{\normalsize Oliver~Fischer\thanks{\texttt{oliver.fischer@liverpool.ac.uk}}}
\author[1,4]{\normalsize A.~Hammad\thanks{\texttt{ahmed.hammad@unibas.ch}}}
\author[3]{\normalsize Christiane~Scherb\thanks{\texttt{cscherb@uni-mainz.de}}}

\affil[1]{\Affilfont Department of Physics, University of Basel, \authorcr 
 		  \Affilfont Klingelbergstr.\ 82, CH-4056 Basel, Switzerland
 		   \authorcr\mbox{}}
\affil[2]{\Affilfont Department of Mathematical Sciences, University of Liverpool
           \authorcr 
 		  \Affilfont  Liverpool, L69 7ZL, UK
 		   \authorcr\mbox{}} 	
 		   	   
\affil[3]{\Affilfont PRISMA$^+$ Cluster of Excellence $\&$ Mainz Institute for Theoretical Physics,
           \authorcr 
 		  \Affilfont  Johannes Gutenberg University, 55099 Mainz, Germany
 		   \authorcr\mbox{}} 		    		   
\affil[4]{\Affilfont Centre for theoretical physics, the British University in Egypt,
           \authorcr 
 		  \Affilfont  P.O. Box 43, Cairo 11837, Egypt
 		   \authorcr\mbox{}} 	
\date{}

\begin{document}
\maketitleNew

\begin{abstract}
\noindent Extra Higgs states appear in various scenarios beyond the current Standard Model of elementary particles. If discovered at the LHC or future colliders, the question will arise whether CP is violated or conserved in the extended scalar sector. An unambiguous probe of (indirect) CP violation would be the observation that one of the extra Higgs particles is an admixture of a CP-even and a CP-odd state. We discuss the possibility to discover scalar CP violation in this way at the high-luminosity (HL) phase of the LHC. We focus on the Two-Higgs Doublet Model of type I, where we investigate its currently allowed parameter region. 
Considering a benchmark point that is compatible with the current constraints but within reach of the HL-LHC, we study the prospects of determining the CP property of an extra neutral Higgs state $H$ via the angular distribution of final states in the decay $H \to \tau\bar\tau$. 
The analysis is performed at the reconstructed level, making use of a Boosted Decision Tree for efficient signal-background separation and a shape analysis for rejecting a purely CP-even or odd nature of $H$.
\end{abstract}

\newpage

\tableofcontents

\section{Introduction}
After the discovery of the scalar resonance with a mass of about 125 GeV, the combined measurements are now used to establish the particle's properties \cite{Khachatryan:2016vau} and whether or not it is indeed the Higgs boson as predicted by the Standard Model (SM). 
An important part of this procedure is the test of its spin~\cite{Choi:2002jk,Gao:2010qx} and CP transformation properties~\cite{DellAquila:1985mtb}, for instance in its top-associated production mode \cite{Buckley:2015vsa}.
The latter is of particular interest, as the violation of the CP symmetry is a fundamental ingredient in order to explain the long-standing puzzle of the matter-antimatter asymmetry of the Universe \cite{Sakharov:1967dj},
in particular because the CP violation in the SM -- observed first in Kaon \cite{Christenson:1964fg} and recently also in Charmed meson decays \cite{Aaij:2019kcg} -- is insufficient.
This calls for physics beyond the SM (BSM) explanations with significant amount of CP violation, which can e.g.\ be introduced in the scalar sector.

According to present analyses, the discovered scalar resonance is compatible with the scalar Higgs boson as predicted by the SM, yet the possibility of a more complex scalar sector that includes CP violation remains.
Although no additional scalar resonances have been found to date\footnote{There exist anomalies in the multi lepton channels and the di-photon channel at the LHC, which were interpreted as possibly due to scalar resonances in refs.~\cite{Abdallah:2014fra,vonBuddenbrock:2019ajh} and \cite{Biekotter:2020cjs}, respectively.} the scalar sector may include additional scalar bosons, which mix only weakly with the SM-like scalar Higgs boson.

A minimal prototype for an extended scalar sector is the Two Higgs Doublet Model (THDM) where the scalar sector of the Standard Model (SM) is extended by an additional scalar $SU(2)_L$ doublet field \cite{Haber:1978jt}, which allows for the possibility of spontaneous violation of the CP symmetry in the scalar sector \cite{Lee:1973iz}, see e.g.\ ref.~\cite{Branco:2011iw} for an overview over its phenomenology.
In general, additional Higgs doublets are tightly constrained as they may introduce Flavour Changing Neutral Currents (FCNCs) at tree-level, and Electric Dipole Moments (EDM) for SM particles, see e.g.\ ref.~\cite{Barger:1989fj}. 

Scalar particles as in the THDM can be discovered and studied at particle colliders, such as the Large Hadron Collider (LHC) \cite{Aad:2015mxa,Khachatryan:2016tnr}.
Once another scalar boson is discovered, its CP properties will be studied, similarly to the Higgs boson, via correlations of the final state leptons from its decays, for instance from sequential gauge boson decays \cite{Chang:1993jy}, polarisation of tau lepton pairs \cite{Berge:2008dr}, or top quark associated production \cite{Gunion:1996xu}.
Recently the state-of-the-art experimental constraints on the type II THDM were combined and it was shown that observable CP-violating effects in the neutron EDM and also in $t\bar t h$ production at the LHC were still possible \cite{Cheung:2020ugr}.

In this paper we go beyond existing studies by investigating in detail the possibility to establish the presence of CP violation in the THDM type I from mixing of heavy neutral scalar particles with different CP transformation properties. 
To this end we define the model in section 2, discuss present experimental constraints and the allowed parameter space in section 3 and perform a collider analysis of the angular distribution of final states in the decay $H \to \tau\bar\tau$ and how it can be used to infer the CP property of extra Higgs states in section 4. We summarise our results and conclude in section 5. In the Appendices A and B we discuss the potential of the alternative decay channel $H \to ZZ \to 4\mu$.

\section{The Complex Two-Higgs Doublet Model}
The THDM was introduced in ref.~\cite{Lee:1973iz} to discuss the phenomenon of CP violation in the scalar sector, an effect that can potentially be large. 
All incarnations of the THDM tend to create tree-level flavor changing neutral currents (FCNCs) that arise from the Yukawa potential.  
In the THDM the FCNCs can be naturally suppressed when a $Z_2$ symmetry is imposed on the Lagrangian \cite{Glashow:1976nt}, as  discussed below.

\subsection{The scalar potential}
In the THDM the scalar sector of the SM is extended by an additional field such that the theory contains two $SU(2)_L$-doublet fields, $\phi_1$ and $\phi_2$, with identical quantum numbers under the SM gauge symmetry group:
\begin{equation}
\phi_1 = \begin{pmatrix}
  \eta_1^+ \\
 (v_1 + h_1 + i h_3)/\sqrt{2} \\
\end{pmatrix}
\hspace{0.5cm} \mbox{and}\hspace{0.5cm}
\phi_2 = \begin{pmatrix}
  \eta_2^+ \\
 (v_2 + h_2 + i h_4)/\sqrt{2} \\
\end{pmatrix}\,.
\end{equation}
Here we introduced the real neutral fields $h_i,\,i=1,...,4$, the charged (complex) fields $\eta_i^+,\,i=1,2$, and the vacuum expectation values (vevs) $v_i,\,i=1,2$.
In its most general form the THDM allows for global transformations which mix these fields and change the relative phases. 
The Lagrangian density for this model can be decomposed as 
\begin{equation}
\mathcal L_{\text{THDM}} = \mathcal L_{\text{SM}, \text{kin}} +  \mathcal L_{\phi, \text{kin}} + V_{\phi} + Y_\phi \,,
\end{equation}
where $ \mathcal L_{\text{SM}, \text{kin}} $ denotes the kinetic terms for SM gauge fields and fermions, $\mathcal L_{\phi, \text{kin}}$ denotes the kinetic terms for the two scalar fields $\phi_i,\,i=1,2$, $V_\phi$ denotes the scalar potential, and $ \mathcal Y_\phi$ the Yukawa terms which gives rise to the couplings between the SM fermions and the scalar fields. 

The most general potential for THDMs can be written as
\begin{equation}\label{eq:1}
\begin{split}
&V_\phi  = m^2_{11} (\phi^\dagger_1 \phi_1) + m^2_{22} (\phi^\dagger_2 \phi_2)-\left[m^2_{12}(\phi^\dagger_1\phi_2)+h.c\right]\\
&\hspace{0.5cm}+ \lambda_1 (\phi^\dagger_1 \phi_1)^2+ \lambda_2 (\phi^\dagger_2 \phi_2)^2 +  \lambda_3 (\phi^\dagger_1 \phi_1) (\phi^\dagger_2 \phi_2)
 +\lambda_4 (\phi^\dagger_1 \phi_2) (\phi^\dagger_2 \phi_1)\\
 & \hspace{0.5cm}+\frac{1}{2}\left[\lambda_5 (\phi^\dagger_1\phi_2)^2+ \lambda_6 (\phi^\dagger_1\phi_1)(\phi^\dagger_1\phi_2)+ \lambda_7 (\phi^\dagger_2\phi_2)(\phi^\dagger_1\phi_2)+H.c.\right]   \,. 
\end{split}
\end{equation}
To avoid FCNCs interactions, THDMs are often defined with a global $Z_2$ symmetry \cite{Glashow:1976nt}, which transforms the scalar fields as
\begin{equation}
\phi_1\to \phi_1 , \hspace{8mm} \phi_2\to - \phi_2\,.
\label{eq:scalar_z2_charge}
\end{equation}
In $V_{\phi}$, this symmetry enforces $\lambda_6=\lambda_7=m^2_{12}=0$. In addition, some of the fermion representations also transform under the symmetry to ensure that only one of the Higgs doublets is involved in each Yukawa matrix. With exact $Z_2$ symmetry, there is no CP violation in the scalar sector, because the only complex parameter in $V_{\phi}$ would be $\lambda_5$, and its effect could be absorbed into global redefinitions of the fields.

To allow for CP violation in the scalar sector of the THDMs, we will consider a softly broken $Z_2$ symmetry, where in addition to $\lambda_5$ also the (complex) parameter $m^2_{12}$ is present and non-zero. The scalar potential is then given by
\begin{equation}
\begin{split}
V_\phi = m^2_{11} (\phi^\dagger_1 \phi_1) + m^2_{22} (\phi^\dagger_2 \phi_2) - \left[m^2_{12}(\phi^\dagger_1\phi_2)+h.c\right]+ \lambda_1 (\phi^\dagger_1 \phi_1)^2+ \lambda_2 (\phi^\dagger_2 \phi_2)^2  \\+\lambda_3 (\phi^\dagger_1 \phi_1) (\phi^\dagger_2 \phi_2) +\lambda_4 (\phi^\dagger_1 \phi_2) (\phi^\dagger_2 \phi_1)  +\frac{1}{2}\left[ \lambda_5 (\phi^\dagger_1\phi_2)^2+H.c\right]  \,.   \hspace{15mm}
\end{split}
\label{eq:Vphi}
\end{equation}
We parametrize the two a priory complex-valued parameters as $ m^2_{12}= |m^2_{12}|e^{i\eta(m^2_{12})}$, $\lambda_5= |\lambda_5|e^{i\eta(\lambda_5)}$, 
introducing the two phases $\eta(m^2_{12})$ and $\eta(\lambda_5)$. 

When minimizing the Higgs potential after electroweak symmetry breaking, the tadpole equations require 
\begin{equation}
\begin{split}
&\frac{\partial V}{\partial h_1} = \frac{1}{2} v_1 v_2^2 \Re(\text{$\lambda_5$})-v_2 \Re(\text{$m_{12}^2 $})+\text{$\lambda_1$} v_1^3+\text{$m^2_{11} $} v_1+\frac{1}{2} \text{$\lambda_3$} v_1 v_2^2+\frac{1}{2} \text{$\lambda_4$} v_1 v_2^2  = 0\:,\\
&\frac{\partial V}{\partial h_2} = \frac{1}{2} v_1^2 v_2 \Re(\text{$\lambda_5$})-v_1 \Re(\text{$m^2_{12} $})+\frac{1}{2} \text{$\lambda_3$} v_1^2 v_2+\frac{1}{2} \text{$\lambda_4$} v_1^2 v_2+\text{$\lambda_2$} v_2^3+\text{$m^2_{22} $} v_2 = 0\:,\\
&\frac{\partial V}{\partial h_3} = -\frac{1}{2} v_1 v_2^2 \Im(\text{$\lambda_5$})+v_2 \Im(\text{$m^2_{12} $}) = 0\:,\hspace{6.cm}\\
&\frac{\partial V}{\partial h_4} = \frac{\partial V}{\partial h_3} \times \left(-\frac{v_1}{v_2}\right) = 0\:,\hspace{8.5cm}
\end{split}
\label{eq:tadpole-equations}
\end{equation}
 with  $v_1$ and $v_2$ denoting the two (by convention real and positive) vacuum expectation values (vevs) of the two scalar fields $\phi_1$ and $\phi_2$. The two vevs satisfy $v = \sqrt{v^2_1+v^2_2}$, with $v$ denoting the SM vev $v \approx 246$ GeV, and we define $\tan\beta := v_2/v_1$. 
 
 Solving the first two equations one can eliminate $m^2_{11}\text{ and } m^2_{22}$ while from the third equation we get the condition $\Im(\text{$m^2_{12} $})=\frac{1}{2} v_1 v_2 \Im(\text{$\lambda_5$})$. In the following we will use this relation to remove $\Im(\text{$m^2_{12} $})$ from all equations, leaving
\begin{equation}
\Re(\text{$m^2_{12}$}) \quad \mbox{and}\quad \lambda_5= |\lambda_5| \,e^{i\eta(\lambda_5)} \,,
\label{eq:complexparameters}
\end{equation}
as the remaining independent parameters. In this sense, the phase parameter $\eta(\lambda_5)$ of $\lambda_5$ governs CP violation in $V_{\phi}$.

\subsection{The mass matrix}
The tree-level mass matrix for the neutral scalars is given by: 
\begin{equation}
({\mathcal{M}}^2)_{ij} = \left. \frac{\partial^2 V}{\partial h_i\partial h_j}\right|_{h_i =0}\,,
\end{equation}
with $h_{i}\ (i = 1,2,3,4)$  being the neutral components of the two Higgs doublets including the Goldstone boson to be absorbed by the Z boson after electroweak symmetry breaking. The mass matrix for the four neutral states in the Higgs basis $h_1,h_2,h_3,h_4$ is

\begin{equation}
M^2 = \left(
\begin{array}{cccc}
D_1 & O_1 & O_2 & O_3 \\
O_1 & D_2 & O_4 & O_5 \\
O_2 & O_4 & D_3 & O_6 \\
O_3 & O_5 & O_6 & D_4
\end{array}\right)\,,
\label{eq:massmatrix}
\end{equation}
with the diagonal elements
\begin{equation}
\begin{array}{ccc}
D_1 & = & 3 \lambda_1 v_1^2+\frac{v_2^2 \lambda_3}{2}+\frac{v_2^2 \lambda_4}{2}+m^2_{11}+\frac{1}{2} v_2^2 \Re(\lambda_5)\,, \\
D_2 & = & \frac{\lambda_3 v_1^2}{2}+\frac{\lambda_4 v_1^2}{2}+\frac{1}{2} \Re(\lambda_5) v_1^2+3 v_2^2 \lambda_2+m^2_{22} )\,,\\
D_3 & = & \lambda_1 v_1^2+\frac{v_2^2 \lambda_3}{2}+\frac{v_2^2 \lambda_4}{2}+m^2_{11}-\frac{1}{2} v_2^2 \Re(\lambda_5)\,, \\
D_4 & = & \frac{\lambda_3 v_1^2}{2}+\frac{\lambda_4 v_1^2}{2}-\frac{1}{2} \Re(\lambda_5) v_1^2+v_2^2 \lambda_2+m^2_{22}\,, \\
\end{array}
\end{equation}
and  the off-diagonal elements
\begin{equation}
\begin{array}{ccc}
O_1 & = & v_1 v_2 \lambda_3+v_1 v_2 \lambda_4+v_1 v_2 \Re(\lambda_5)-\Re(\text{$m^2_{12}$})\,, \\
O_2 & = &  -\frac{1}{2} v_2^2 \Im(\lambda_5)\,, \\
O_3 & = & \frac{1}{2} v_1 v_2 \Im(\lambda_5)\,,  \\
O_4 & = & -O_3 \,,\\ 
O_5 & = &  \frac{1}{2} v_1^2 \Im(\lambda_5)\,, \\
O_6 & = & v_1 v_2 \Re(\lambda_5)-\Re(\text{$m^2_{12}$})\,.\\
\end{array}
\end{equation}
Diagonalizing the mass matrix in eq.~\eqref{eq:massmatrix} leads to three massive neutral scalar bosons $H_1,H_2$ and $H_3$, and one massless neutral field $H_0$.
In this article we will evaluate the mass matrix numerically. 
An analytical dependence of the mass eigenstates' physical properties on the model parameters can be extracted under certain simplifying assumptions, see e.g.\ refs.~\cite{Ginzburg:2004vp,Fontes:2017zfn}. 

In general, the mass eigenstates do not conserve the CP symmetry. One can see that with  the only source of CP violation coming from  $\Im(\lambda_5)$, for  $\Im(\lambda_5)\to 0$ one retains the CP conserving THDM (with vanishing off-diagonal entries in the mass matrix, $O_{2,3,4,5} \to 0)$.
The squared neutral Higgs mass matrix can be diagonalized by a $4\times 4$ matrix $R$ as 
\begin{equation}
R^\dagger {\mathcal{M}}^2 R = {\mathcal{M}}^2_{diag} =\mbox{diag}(0,M^2_{H_1},M^2_{H_2},M^2_{H_3})\,.
\label{eq:rotationmatrix}
\end{equation}
The neutral Higgs mass eigenstates $H_i\ (i=0,1,2,3)$ are related to the interaction fields $h_i$ via the rotation 
\begin{equation}
h_i = \sum_i R_{ij} H_j\,.
\end{equation}
In the following we identify $H_0$ with the Goldstone boson that is absorbed by the $Z$ boson and $H_1$ with the SM-Higgs-like scalar resonance at $\sim$ 125 GeV.
This leaves the neutral bosons $H_2$ and $H_3$ as new scalar mass eigenstates yet to be observed. We will assume that the extra Higgs states are heavier than $H_1$ and, without loss of generality, require the mass ordering $M_{H_1}\le M_{H_2}\le M_{H_3}$. 
The evaluation of the mass matrix and the rotation matrix $R$ is carried out numerically using SPheno\cite{Porod:2003um, Porod:2011nf}.

\subsection{The Yukawa sector}
The absence of FCNCs at tree-level is ensured when a basis exists in which the contributions to the mass matrices for each fermion of a given representation stem from a single source \cite{Glashow:1976nt,Paschos:1976ay}. 
In the Standard Model with left-handed doublets and right-handed singlets, this implies that all right-handed quarks of a given charge must couple to a single Higgs multiplet, which can be ensured via a discrete $Z_2$ symmetry. 

This symmetry transforms the scalar fields as in eq.~\eqref{eq:scalar_z2_charge}, and allows for different possible $Z_2$ charge assignments for the SM fermions.
Here, we select the $Z_2$ charge assignment of the type I version of the THDM, where all quarks and charged leptons couple only to one of the scalar doublet fields, conventionally chosen to be  $\phi_2$.\footnote{In the THDM model of type II, the up-type (down-type) quarks and leptons couple conventionally only to  $\phi_2$ ($\phi_1$). 
Further variations in the lepton sector exist: the ``lepton specific'' model, where all quarks couple to $\phi_2$ while the leptons couple to $\phi_1$, and the ``flipped'' model, where right-handed leptons couple to $\phi_2$ like the up-type quarks \cite{Branco:2011iw}. }

The $Z_2$-symmetric Yukawa terms of type I THDM are given by 
\begin{equation}
-\mathcal{Y}_\phi =  Y_u\bar{Q}_L i\sigma_2\phi^\ast_2 u_R  + Y_d\bar{Q}_L \phi_2 d_R + Y_e\bar{L}_L \phi_2 e_R + \mbox{H.c.}	
\end{equation}
with the Yukawa coupling matrices $Y_u,\,Y_d,\,Y_e$.

\subsection{CP violation}
\label{subsec:CPviolation}
The scalar potential of eq.~\eqref{eq:Vphi} in general mixes the interaction states with definite CP transformation properties.
This is clearly visible in the mass matrix of eq.~\eqref{eq:massmatrix}, which mixes the CP-even $h_1,h_2$ with the CP-odd $h_3,h_4$ when at least one of the off-diagonal entries $O_i,\,i=2,3,4,5$ is non-zero, i.e. when $\lambda_5$ has a non-zero imaginary part. The proposed methods for testing CP violation in the Higgs sector include:
\begin{itemize}
	\item If an extra Higgs state $H_i$ is discovered, its top quark associated production cross section could be used to determine its CP property~\cite{Gunion:1996xu,Boudjema:2015nda,Goncalves:2016qhh,AmorDosSantos:2017ayi}, because it is sensitive to the relative magnitudes of the CP-even and CP-odd coefficients of the $\bar t t H_i$ coupling. However, this effect is suppressed by the smaller cross section of a three particle final state.
	\item The angular momentum correlations of the final state muons in $H_i\to ZZ\to 4 \mu$ have been proposed as a method to determine the CP transformation property of $H_i$~\cite{Choi:2002jk,Buszello:2002uu,Godbole:2007cn,Chang:1993jy,Ferreira:2016jea,DeRujula:2010ys,Bolognesi:2012mm,Artoisenet:2013puc}. We will discuss the applicability of this method in the context of the THDM of type I in Appendix~\ref{appendix_A}. We find that at the HL-LHC the loop-induced decay rate of the CP-odd pseudoscalar (or of the CP-odd component of a mixed state) via $ZZ$ into $4 \mu$ is too suppressed for successful application of the method.
	\item When contributions from loop-level decays of the $H_i$ can be neglected, an obvious sign for CP mixing in the THDM is the simultaneous observation of three different Higgs states with interactions that, in the CP conserving case, are only possible at tree-level for pure CP eigenstates \cite{Fontes:2015xva}.
One example is the scalar decay chain $H_i\to ZZ\to 4 \mu$ mentioned above. Since $H_i \to ZZ$ at tree-level is only possible when the $H_i$ has a CP-even component, in the CP conserving THDM only $H_1$ and either $H_2$ or $H_3$ can have this tree-level decay. Observing it for all three $H_i$ one can conclude that the THDM violates CP. We discuss this example in Appendix \ref{appendix_B}.
However, it is important to note that the observation of several scalar resonances with decays into $ZZ$ is not an unambiguous signal of CP mixing in general, since the third resonance could stem from additional scalar fields outside the THDM.
	\item The CP transformation property of the $H_i$ can be inferred from its decays to tau lepton pairs.
	To be specific, the correlation of the tau lepton polarisation planes are directly linked to the CP properties of the parent, and they can be reconstructed via the hadronic decay modes of the two tau leptons \cite{Berge:2008dr,Harnik:2013aja,Berge:2015nua,Bhardwaj:2016lcu,Lasocha:2020ctd,Swain:2020sil}. In the following we will focus on this method in the main part of the paper.
\end{itemize}

\subsection{Discovering CP violation via $H\to\tau\bar\tau$}
\label{sec:discoveringCPviatautau}
We use the impact parameter method as first presented in ref.~\cite{Berge:2008dr} to extract transverse spin correlations in the decay chains of a field $S$ which is a mixture of a scalar and a pseudoscalar field. 
In particular we focus on decay chains of the form $S\to\tau\bar\tau$ with $\tau^\pm\to\pi^\pm\ \bar{\nu}_\tau(\nu_\tau)$ and make use of the impact parameters of the visible decay products of the tau lepton, $\tau_{\mathrm{had}}$, to extract an asymmetry in the acoplanarity angle of the two tau leptons.
We remark that the method does not depend on the Higgs boson production mechanism, but translates directly into correlations among their decay products.

The Yukawa interaction of S can be written as 
\begin{align}
{\mathcal{L}_y} = y_{S \tau}\, \left(\bar{\tau}\left(C_v + C_a \ i \gamma_5\right) \tau\right) S
\end{align}
with $y_{S\tau}$ being the effective Yukawa coupling of $S$ and the tau lepton
and $C_a,\,C_V$ being the scalar and pseudoscalar components of the coupling, respectively, with $C_v^2+C_a^2=1$. The effective mixing angle $\theta_{\tau\tau}$, defined as
\begin{align*}
\text{tan}(\theta_{\tau\tau}) = \frac{C_a}{C_v}\,,
\end{align*} 
measures the mixing of CP eigenstates. For example, $\theta_{\tau\tau}= 0\ (\frac{\pi}{2})$ holds for pure scalar (pseudoscalar) coupling.

The $\tau\bar\tau$ spin correlation can be inferred from the angle between the tau decay planes. 
We remark that we  consider here only the tau decay mode $\tau^\pm\to\pi^\pm\bar{\nu}_\tau(\nu_\tau)$, which has a branching fraction of $ 11\%$.
While this limits our statistics it provides a clear signal and can thus serve as a conservative estimate for the sensitivity to distinguish CP properties.

For the tau decay $\tau^\pm\to\pi^\pm\bar{\nu_\tau}(\nu_\tau)$ the angular correlation in the decay width can be written as \cite{Berge:2008dr}:
\begin{align}
\frac{1}{\Gamma} \frac{d\Gamma}{d\phi}  = \frac{1}{2\pi}\left(1-\frac{\pi^2}{16} \frac{C_v^2-C_a^2}{C_v^2+C_a^2}\cos\phi\right)= \frac{1}{2\pi}\left(1-\frac{\pi^2}{16} \left(\cos^2\theta_{\tau\tau}-\sin^2\theta_{\tau\tau}\right)\cos\phi\right)\,.
\label{eq:theorydistribution}
\end{align}
The angle $\phi$ between the decay planes is the so-called acoplanarity angle, which is sensitive to the CP properties of the scalar parent $S$ via the coupling parameters $C_v$ and $C_a$. The angular correlation in eq.~\ref{eq:theorydistribution} is given for the case in which one cannot distinguish $\phi$ from $2\pi-\phi$ and is obtained by the sum over both cases \cite{Berge:2008dr}. The acoplanarity angle ($\phi$) can be reconstructed from the tau decay properties, namely the two impact parameter vectors
\begin{equation}
\phi = \arccos(\vec{n}^- \cdot \vec{n}^+)\,,
\end{equation} 
where we introduced
\begin{align}
\vec{n}^\pm = \frac{\vec{P}_{\pi^\pm}\times \vec{P}_{\tau^-}}{|\vec{P}_{\pi^\pm}\times \vec{P}_{\tau^-}|}\,.
\end{align}
The impact parameter is defined as the shortest path between the primary vertex and the pion momentum vector extended in the direction of the tau decay point. 
Since it is basically impossible to reconstruct the tau lepton momentum due to the presence of tau neutrinos among the decay products, the authors in ref.~\cite{Berge:2008dr} introduce the so-called ``Zero-Momentum-Frame'' (ZMF) of the tau decay products, in our case the pions.
This does not affect the correlations of the decay planes, such that the exact tau direction does not matter.
We find the ZMF by boosting the meson momenta such, that $\vec P_{\pi^+}^\ast = -\vec P_{\pi^-}^\ast$, where quantities with an asterisk ($^\ast$) refer to the ZMF.
Then a 4-vector is defined for the normalized impact parameter for each tau lepton in the ZMF as $n^{\ast\pm} = (0,\vec{n}^{\ast\pm})$, from which one can extract the acoplanarity angle in the boosted frame:
\begin{align}
\phi^\ast = \arccos(\vec{n}^{\ast -}_\bot \cdot \vec{n}^{\ast +}_\bot)\,.
\label{eq:acoplanarityZMF}
\end{align}
The resulting distribution for $\phi^\ast$ between 0 and $\pi$ allows for a clear distinction of fields that are even or odd eigenstates of CP. 
Below, in sec.~\ref{sec:probingCPviolation}, we will analyse how well the CP property of an extra Higgs state can be distinguished using this method. Finally, we note that the distribution in eq.~\eqref{eq:theorydistribution} remains invariant when switching from $\tau\bar\tau$ in the laboratory frame to $\pi^+\pi^-$ in the ZMF (cf. also ref.~\cite{Berge:2008dr}):
\begin{align}
	\frac{1}{\Gamma} \frac{d\Gamma}{d\phi^\ast}  =\left.\frac{1}{\Gamma} \frac{d\Gamma}{d\phi} \right|_{\phi\to \phi^\ast} =\frac{1}{2\pi}\left(1-\frac{\pi^2}{16} \left(\cos^2\theta_{\tau\tau}-\sin^2\theta_{\tau\tau}\right)\cos\phi^\ast\right)\,.
	\label{eq:theory_prediction}
\end{align}

\section{Constraints}
The THDM with CP violation is constrained from various observations and measurements at collider and non-collider experiments.
Below, we discuss constraints from theoretical considerations, from $B$-physics measurements, Higgs data (from the LHC, LEP, and the Tevatron) and from measurements of EDMs.

\subsection{Theory considerations}
As first condition from theory we impose that it has to be perturbative, which constrains the magnitude of the couplings $|\lambda_i|\lesssim 4\pi$.
The second theory condition that each model has to satisfy is the stability of the vacuum. Therefore, the potential should be positive for large values of $\phi$, which leads to the constraints \cite{Ginzburg:2004vp,Arhrib:2010ju}:
\begin{equation}
\lambda_1 > 0,\hspace{8 mm} \lambda_2>0,\hspace{8 mm} \lambda_3 + \sqrt{\lambda_1 \lambda_2}>0,\hspace{8 mm}\lambda_3+\lambda_4-|\lambda_5|+\sqrt{\lambda_1 \lambda_2}>0\:.
\end{equation}
The third condition is that the S-matrix has to be unitary for an elastic two-to-two boson scattering process, which limits the magnitude of $\lambda_i$, cf.\ refs.~\cite{Ginzburg:2004vp,Arhrib:2010ju}. 
The fourth condition stems from the so-called oblique parameters, which are constrained as (cf.\ the global fit of \cite{Baak:2012kk,Baak:2011ze}):
\begin{equation}
S = 0.03 \pm 0.10 ,\hspace{10 mm} T = 0.05 \pm 0.12,\hspace{10 mm} U = 0.03 \pm 0.10\,.
\end{equation}
These parameters receive contributions from the THDM at the loop-level, and  present an independent important constraint.

\subsection{B physics data}
The charged Higgs bosons from the THDM contribute to the decays of $B$ mesons, such that the $B$-physics data set can be used to constrain the THDM parameters.
Since the couplings of the charged Higgs bosons are not sensitive to the parameters of the neutral sector, these constraints are independent of the amount of CP violation in the model. 

To evaluate the flavor phenomenology in particular for the $B$ physics processes we use the numerical tool FlavorKit~\cite{Porod:2014xia}, which evaluates many flavor-related observables for every scanned point.
The most stringent constraints on our model parameters stem from the process $B\to X_s\gamma$, which limits in particular the charged Higgs mass: $m_H^\pm \ge 580$ GeV at $\tan\beta = 1$ for the THDM of type II. For the type I THDM, the strongest constraints on the charged Higgs mass apply for $\tan\beta\le 2$, while with increasing $\tan\beta$ the constraints get weaker, see refs.~\cite{Misiak:2017bgg,Enomoto:2015wbn}.
We use the experimental bounds as reported in \cite{Amhis:2019ckw}:
\begin{equation}
\mbox{Br}(B\to S\gamma)_{E_\gamma \ge 1.6 \,\mathrm{GeV}} \leq (3.32\pm 0.15) \times 10^{-4} \:.
\end{equation}

\subsection{Higgs data}
The global data set on the Higgs boson includes results from LEP, the Tevatron and the LHC experiments.
The existing data is combined with the numerical tool HiggsBounds\cite{Bechtle:2013wla}, which we employ to constrain the THDM parameter space. 

HiggsBounds first identifies the most sensitive signal channel for each boson $H_i$ separately and then computes the ratio of this theoretically predicted to the observed signal strength for heavy Higgs bosons as
\begin{equation}
{\mathcal{K}_i}= \frac{\sigma\times \mbox{Br}(H_i)_{model}}{\sigma\times \mbox{Br}(H_i)_{obs}}\,,  
\end{equation}
which we use to obtain an exclusion limit at $95\%$ C.L for parameter space points where at least one observable exists, such that ${\mathcal{K}_i}> 1$.

In addition to the exclusion of individual parameter points, we employ the numerical tool HiggsSignals \cite{Bechtle:2013xfa} to evaluate the statistical compatibility of the lightest SM-like Higgs boson in the model with the observed scalar resonance, as it is observed by the LHC experiments.
Also, the SM-like Higgs signal rates and masses are compared with the various signal rate measurements published by the experimental collaborations for a fixed Higgs mass hypothesis. 
The model is tested at the mass position of the observed Higgs peak in the channels with high mass resolutions like $h\to ZZ^\ast\to 4\ell$ and $h\to\gamma\gamma$. The signal strength modifier for the model for one channel is calculated  as 
\begin{equation}
\mu = \frac{(\sigma\times \mbox{Br})_{model}}{(\sigma\times \mbox{Br})_{SM}}\times \omega\,,
\end{equation}
with $\omega$ being the SM weight, including the experimental efficiency. 

A $\chi^2$ test for the model hypothesis is performed, where a local excess in the observed data at a specified mass is matched by the model. The  signal strength modifiers and the corresponding predicted Higgs masses enters the total $\chi^2$ evaluation as 
\begin{equation}
\chi^2_{tot} = \chi^2_\mu + \sum_{i=1}^{N_H} \chi^2_{m_{H_i}}\,,
\end{equation}
where $\chi^2_\mu $ is the $\chi$-squared measure calculated from the signal strength modifier only and $\chi^2_{m_{H_i}}$ is the $\chi$-squared measure calculated from Higgs bosons mass, with $i$ running over the number of the neutral Higgs bosons in the model.  The intrinsic experimental statistical and systematic uncertainties within $1\sigma$ for $\chi^2_\mu $  is given by
\begin{equation}
\chi^2_\mu = (\mu_{obs}-\mu_{model})^T C^{-1}_{ij}(\mu_{obs}-\mu_{model})\,,
\end{equation}
where $C_{ij}$ is the signal strength covariance  matrix that contains the uncorrelated intrinsic experimental statistical and systematic uncertainties  in its diagonal entries.

The $1\sigma$ and $2\sigma$ error can be obtained from the best-fit value as $1(2)\sigma =\Delta\chi^2_{best} +2.3(5.9) $ with $\Delta\chi^2_{best} = 1.049$. CMS reports the combined best fit value for the SM Higgs signal strength at center of mass energy = $13$ TeV and integrated luminosity = $35.9 fb^{-1}$ to be $\mu_{best} =1.17^{+ 0.1}_{- 0.1}$\cite{CMS:2018lkl}, while the recent ATLAS results at $\sqrt{S}=13$ TeV and integrated luminosity = $79.8 fb^{-1}$ reports $\mu_{best} =1.13^{+ 0.09}_{- 0.08}$\cite{ATLAS:2018doi}.
These results put strong constraints on the physical properties of $H_1$ to be close to the ones of the SM Higgs boson.
It also limits strongly the possible amount of mixing between $H_1$ and $H_i,\,i=2,3$.
\subsection{Electric Dipole Moments}
\begin{figure}
\centering
\includegraphics[scale=0.25]{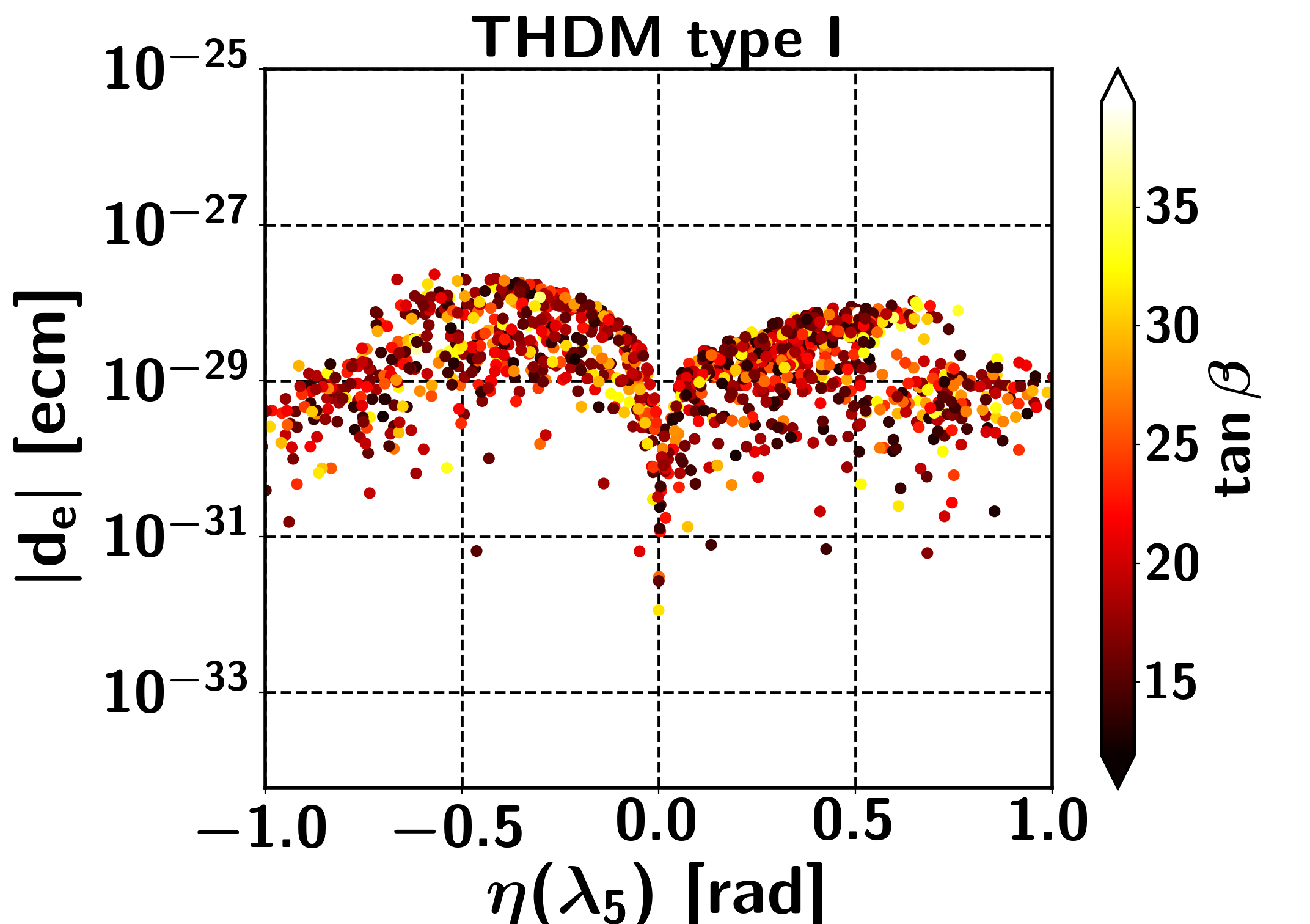}
\quad
\includegraphics[scale=0.25]{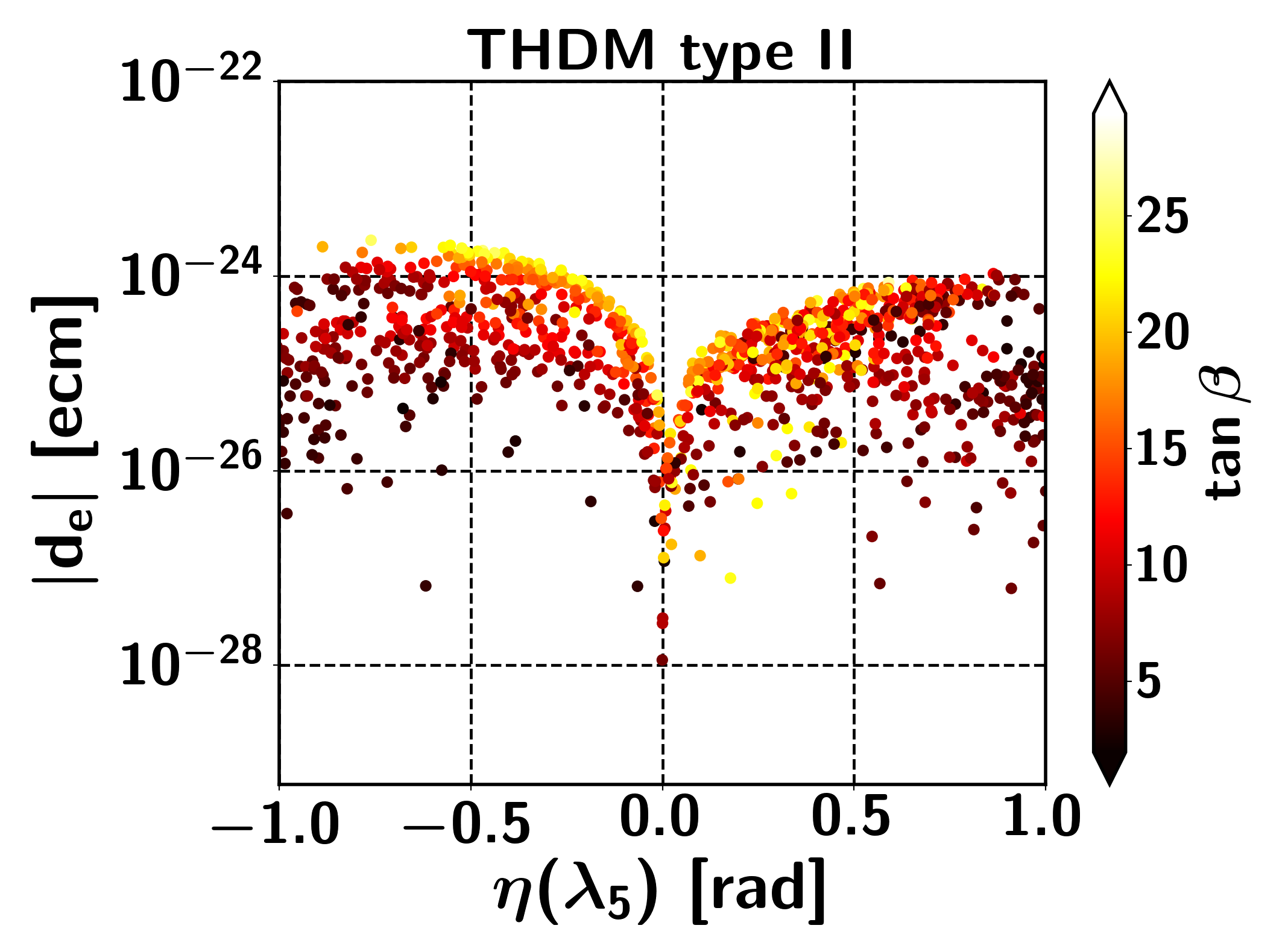}	
	\caption{Electron EDM versus $\eta(\lambda_5)$ as a function of $\tan\beta$ for type I and type II THDMs. 
	Points in the plots satisfy all constraints including the Higgs data (at $2\sigma$). }
	\label{fig:de_comparsion}
\end{figure}
The upper limit on the electric dipole moment (EDM) of the electron is $|d_e|<1.1\times10^{-29}$~ecm \cite{Andreev:2018ayy}. The new scalars contribute to the electron EDM via Barr-Zee diagrams as discussed, e.g., in refs.~\cite{Abe:2013qla} and \cite{Chun:2019oix} (for latest two-loop results, see ref.~\cite{Altmannshofer:2020shb}).
In particular, the CP violating complex phase is found to strongly affect the magnitude of the EDM, its main source being the modified couplings of the Higgs bosons. In the type II THDM the EDM is enhanced by $\tan\beta$, while in type I the EDM is suppressed by $\frac{1}{\tan\beta}$, cf.\ ref.~\cite{Egana-Ugrinovic:2018fpy}. 

As stated above, the Yukawa couplings can be expressed as a sum of their CP-even and CP-odd part. 
In general, if a fermion couples to $\phi_1$ ($\phi_2$) both parts of the coupling are proportional to $\tan\beta$ ($\frac{1}{\tan\beta}$). 
Thus, in the type I THDM all Yukawa couplings are proportional to $\frac{1}{\tan\beta}$, while in the type II THDM the Yukawa couplings of down-type quarks and leptons are proportional to $\tan\beta$.

In this article we consider large $\tan\beta$, which leads to potentially large couplings and large contributions to the EDM. 
Therefore, for large $\tan\beta$ the type I THDM with couplings proportional to $\frac{1}{\tan\beta}$ is less constrained, which is one reason for us to focus on this version of THDM.
To analyse the EDM constraint, we employ the formulae from refs.~\cite{Abe:2013qla,Chun:2019oix}.

\subsection{Scanning the parameter space}
\label{sec:scan}
In order to find viable parameter space points that satisfy all constraints we perform a scan over the parameter space.
In this scan the full parametric dependence of the physical properties of the scalar particles, like their masses and interaction vertices, are calculated, and the above described constraints are evaluated.
For the numerical scan we consider the following ranges of parameters:
\begin{equation}
\begin{split}
0.0 \le \lambda_1\le10,\hspace{5 mm} 0.05 \le \lambda_2\le0.2,\hspace{5 mm} 0\le \lambda_3\le 10,\hspace{5 mm} -10 \le \lambda_4\le 10,\hspace{2cm} \\ -10 \le |\lambda_5|\le 10, 
-1.0 \le \eta(\lambda_5)\le 1.0,\hspace{5 mm} 2 \le \tan\beta\le 50,\hspace{5 mm} -25 \text{ TeV}^2 \le m^2_{12}\le 25 \text{ TeV}^2.
\end{split}
\label{eq:parameter_ranges}
\end{equation}
We obtain 5k parameter space points that satisfy all experimental constraints.
We remark that the above parameter ranges are optimised to yield a good efficiency with respect to passing the list of constraints.
As we mentioned above we use SPheno to evaluate the mixing matrix numerically.

In fig.~\ref{fig:de_comparsion} we show the contribution to the EDM for our parameter space points as a function of $\tan\beta$ and $\eta(\lambda_5)$ for the type I THDM (left panel). We also show the results for the type II THDM for comparison in the right panel of the same figure.
One can see that for the THDM of type II low $\tan\beta$ with large $\eta$ has the smallest EDM values. 
With the used scan resolution no points below the EDM bound are found. 
This can be compared with the analysis in ref.~\cite{Cheung:2020ugr}, wherein a region with small values for $\tan\beta$ was identified that is not excluded by the EDM and the Higgs constraints considering the type II THDM. 
For the type I version of the THDM allowed parameter space points can be found for all considered values of $\eta(\lambda_5)$.
\begin{figure}
\centering
\includegraphics[scale=0.25]{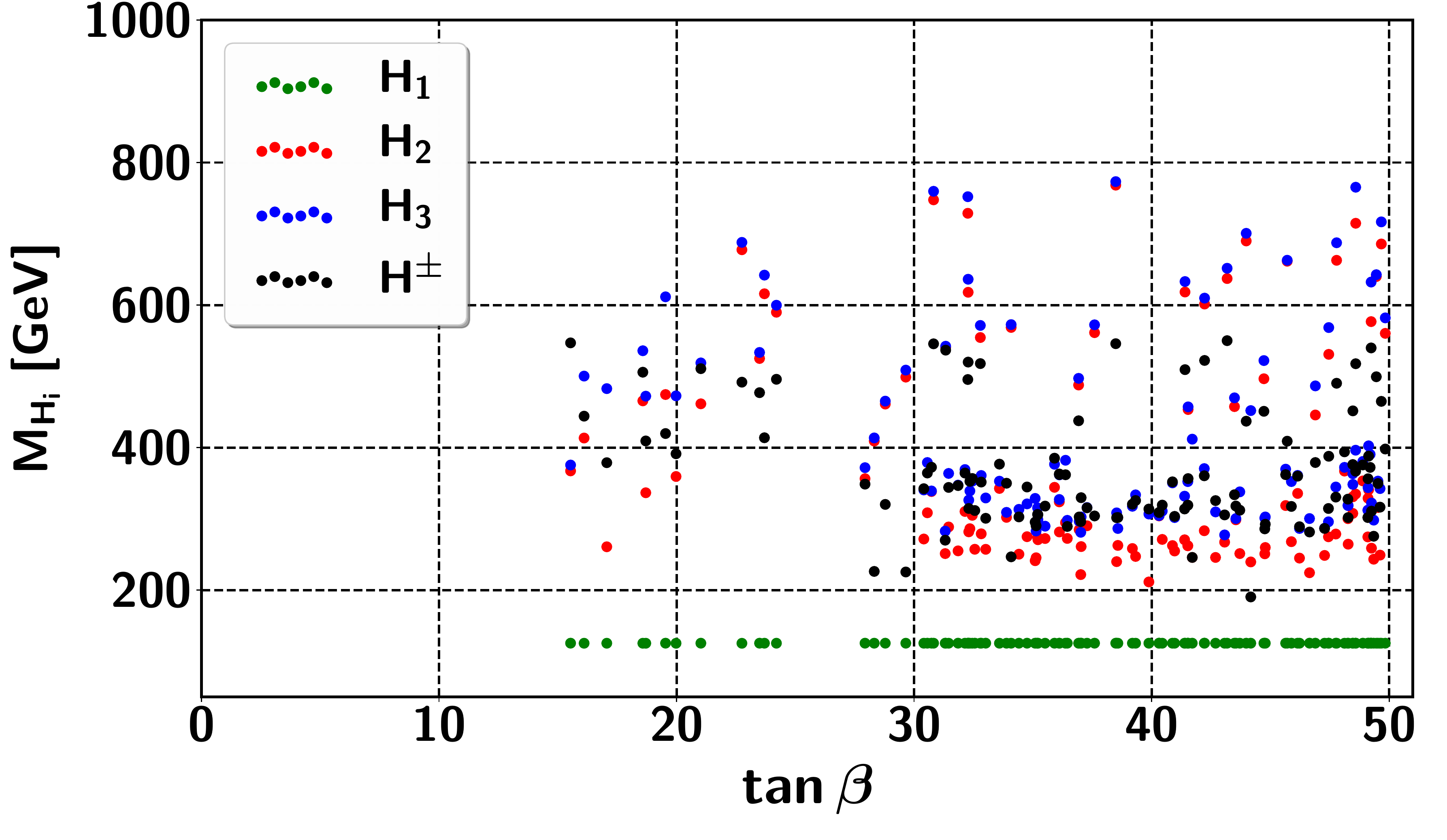}
\caption{Scatterplot of the allowed parameter space points in the projection of mass $m_{H_i}$ (in GeV) over $\tan\beta$.}
\label{fig:mass_tanbeta}
\end{figure}
For the parameter space points satisfying all of the above constraints we show the projection of the three neutral scalar masses versus $\tan\beta$ in fig.~\ref{fig:mass_tanbeta}.
From this figure we can see that for the viable points we found in our scan, both, $H_2$ and $H_3$, have masses between about 200 and 700 GeV, while $H_2$ has more parameter space points with masses around 200 to 300 GeV, and $H_3$ tends to be slightly heavier.

\section{Analysis}
\label{sec:probingCPviolation}
In this section we discuss the production mechanism for the scalar bosons of the type I THDM and the currently allowed cross sections.
We investigate the process $pp\to H_i \to \tau\bar\tau$ that we use to analyse the CP properties of extra Higgs states and evaluate the prospects of finding it in the presence of  the considered background processes. Then we perform an analysis of the angular distribution of the final state taus.
\subsection{Heavy scalar production rates}

We consider the LHC in its high-luminosity phase (the HL-LHC) with an expected total integrated luminosity of 3 ab$^{-1}$ and a center-of-mass energy of 14 TeV.
The dominant production processes for the Higgs bosons at the HL-LHC are gluon-gluon fusion (around $90\%$) and vector boson fusion. We calculate the effective gluon-gluon-Higgs coupling using SPheno and include the QCD corrections from ref.~\cite{Belyaev:2012qa}. The production cross sections are calculated  including the effective gluon-Higgs vertex in MadGraph~\cite{Alwall:2014hca}.

\begin{figure}
\centering
\includegraphics[scale=0.25]{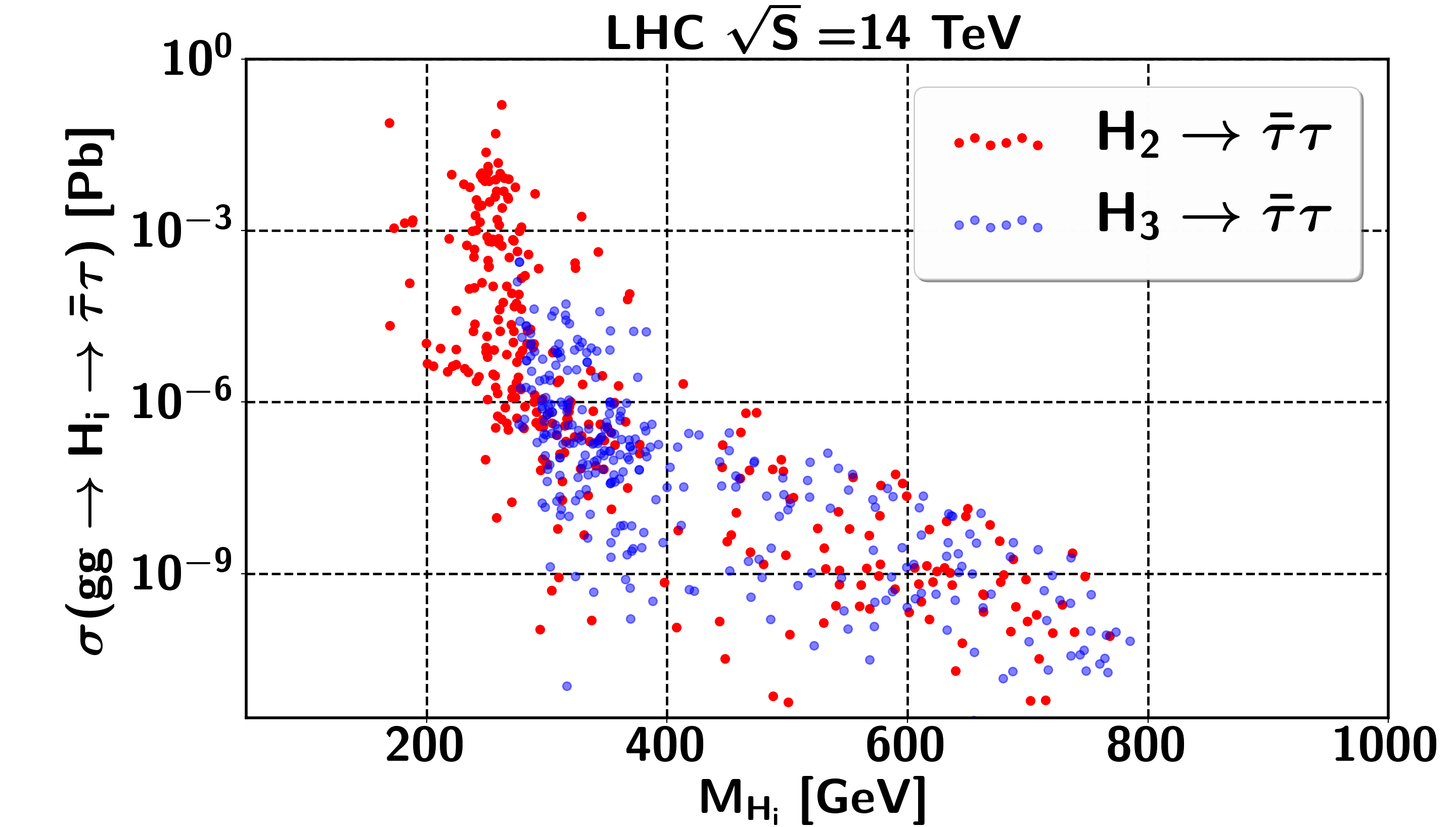}\\
\caption{Total cross sections for the process $pp\to H_i\to\tau\bar{\tau}$ at the HL-LHC with $\sqrt{s}=14$ TeV.}
\label{fig:6}
\end{figure}

Since the signal for CP violation is encoded in angular correlations of the heavy scalars' decay products, it can only be assessed statistically. 
Therefore we are interested in how many signal events can be expected, requiring that the parameter point is allowed by the above discussed constraints.
For this assessment, we use our parameter space set from the previous section, selecting for parameter points that conform with all constraints.
We show the total cross section for the process $pp \to H_i\to\tau\bar{\tau}$ in fig.~\ref{fig:6}, wherein the blue and red points denote the cross sections for the scalar bosons $H_2$ and $H_3$, respectively.

We notice that parameter space points exist with production cross sections larger than a few femtobarn, which would yield more than a few thousand events at the HL-LHC.
While this is in principle sufficient for a statistical study of the CP violation signal, it may be difficult in practice due to large backgrounds and reconstruction uncertainties.
In the next subsection, we will evaluate a specific benchmark point.
\subsection{Signal reconstruction for a benchmark point at the HL-LHC}
\label{sec:reconstruction_tau}
In the following we discuss the inclusive signal process
\begin{equation}
\qquad pp \to H_i \to \tau\bar\tau\,,
\end{equation}
where we include interference between the $H_i$.
We select a benchmark point with $m_{H_2} = 250$ GeV and $m_{H_3} = 300$ GeV, based on the model parameters $\tan\beta=31$, $\theta_{\tau\tau} = 0.68 = \frac{\pi}{4.6}$ (which corresponds to $\eta(\lambda_5) = 0.7$), $\lambda_1=0.039,\ \lambda_2=0.104,\ \lambda_3=2.215,\ \lambda_4=-0.023,\  \Re(\lambda_5)= 0.337$ and $m_{12}^2=-1.919\times10^{4}$ GeV$^2$.
The parameters $m^2_{11}$ and $m^2_{22}$ are then fixed by the previous parameters due to the tadpole equations, cf.\ eq.~\eqref{eq:tadpole-equations}. It is worth noting that the benchmark point is stable against small changes in the input parameters, e.g changes in the input parameters of $\mathcal{O}(5\%)$ lead to changes in the masses of $\mathcal{O}(0.1\%)$ while still fulfilling all above discussed constraints. Our benchmark point has an electron EDM $|d_e|\approx 7.4\times10^{-30}$~ecm and a branching ratio $Br(B\to X_s\gamma)\approx 3.04\times10^{-4}$ leading to possible observable signatures. Therefore both channels can be used as complementary probes of our benchmark point. 

The main irreducible SM backgrounds to this process come from $Z\to\tau\bar\tau$\cite{Aad:2015vsa} and from single top and $\bar{t} t$, with tau jet pair produced from the $W$ decay.
Other backgrounds arise from the misidentification of light jets as tau jets, for instance $W$ boson plus jet or multijets.
The here considered backgrounds are listed, together with their cross sections, in tab.~\ref{tab:backgrounds}
\begin{table}
\begin{center}
\begin{tabular}{|l|c|}
\hline
Backgrounds & $\sigma_{(HL\text{-}LHC)} [Pb]$   \\ \hline\hline
$Z\to\tau\bar\tau$  & 1537   \\ \hline
QCD jets & $10^8 \times \epsilon^2$ \\ \hline
$W+J, W\to\tau\ \nu_\tau$  & 22  \\ \hline
$t\bar t $  &  6 \\ \hline
$WW, W\to\tau\ \nu_\tau$  & 0.9  \\ \hline
\end{tabular}
\end{center}
\caption{Dominant background processes considered in our analysis and their total cross sections. The samples have been produced with the following cuts: $P_T(j)\ge 20$ GeV, $P_T(l)\ge 10$ GeV. The efficiency of the QCD jets to be mistagged as tau jet is taken from the CMS paper\cite{CMS:2020rpr} and we use the fake rate $\epsilon = 5\times 10^{-3}$ from ref.~\cite{CMS:2020rpr}.}
\label{tab:backgrounds}
\end{table}

We simulate signal samples including $20$ million  events and background samples including $30$ million events for each background with MadGraph5\cite{Alwall:2014hca}. The parton shower, hadronisation and spin correlation of the tau lepton decay is taken care of by Pythia8 \cite{Sjostrand:2007gs}. 
We perform a fast detector simulation with Delphes \cite{deFavereau:2013fsa}.
The tau jets are tagged using the Delphes analysis framework with reconstruction efficiency of $70\%$ and misidentification rate of $5\times10^{-3}$ for the QCD jet, which we implement at the analysis level.
For the background we adopt a reconstruction efficiency of 60\% (following ref.~\cite{Bagliesi:2007qx}). 
For the event reconstruction we require two tau tagged jets with $P_T > 20$ GeV where events with $b$-tagged jets are rejected. 

We find that interference between the $H_i$ bosons has a very small effect for the here chosen benchmark point, namely it increases the total cross section by about 5\%. In particular, the interference between $H_2$ and $H_3$ is suppressed by the small $H_3$ total cross section, which is about $1.5 \cdot 10^{-5}$ pb, compared to the total cross section of the $H_2$, which is $0.3$ pb.
Therefore, in the next section, we will study an exclusive sample from the process $pp \to H_2 \to \tau\bar\tau$.

\subsection{Shape analysis for establishing CP violation}
\begin{figure}
	\centering
	\includegraphics[width=7cm,height=4.8cm]{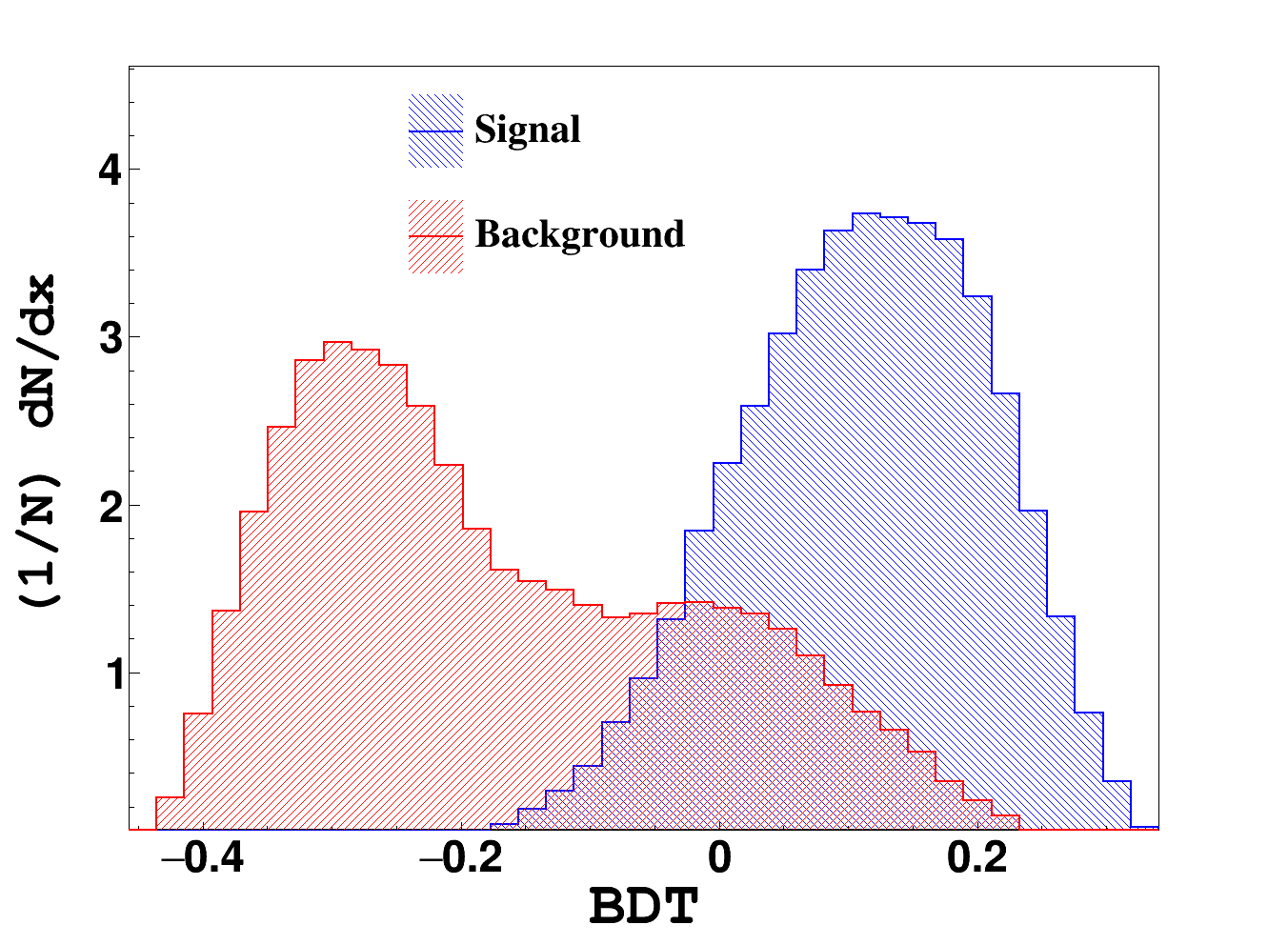}\includegraphics[width=8cm,height=5cm]{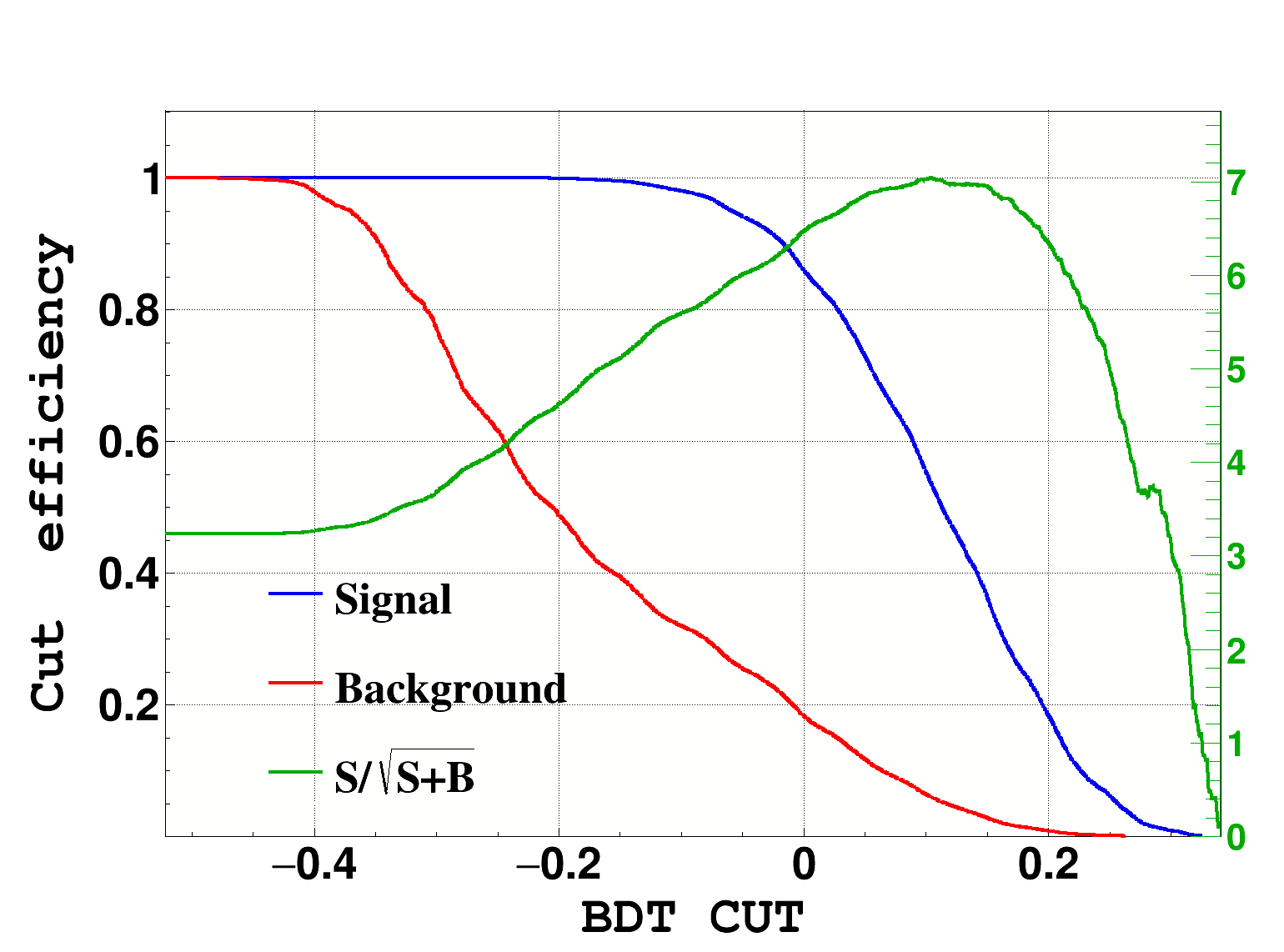}
	
	\caption{Left: The distribution of the Boosted Decision Tree response to the signal (blue) and to the background (red), superimposed. Right: Cut efficiency that maximizes the BDT cut. For a cut value greater than 0.104 one can
		get $\frac{S}{\sqrt{S + B}} = 7.04\sigma$ with number of signal events = 2043 and background events = 82212 after the BDT cuts. The cut efficiency for the signal is 0.57 and for the background 0.00059.}
	\label{fig:8}
\end{figure}

\begin{figure}[h!]
\centering
\includegraphics[width=0.9\textwidth]{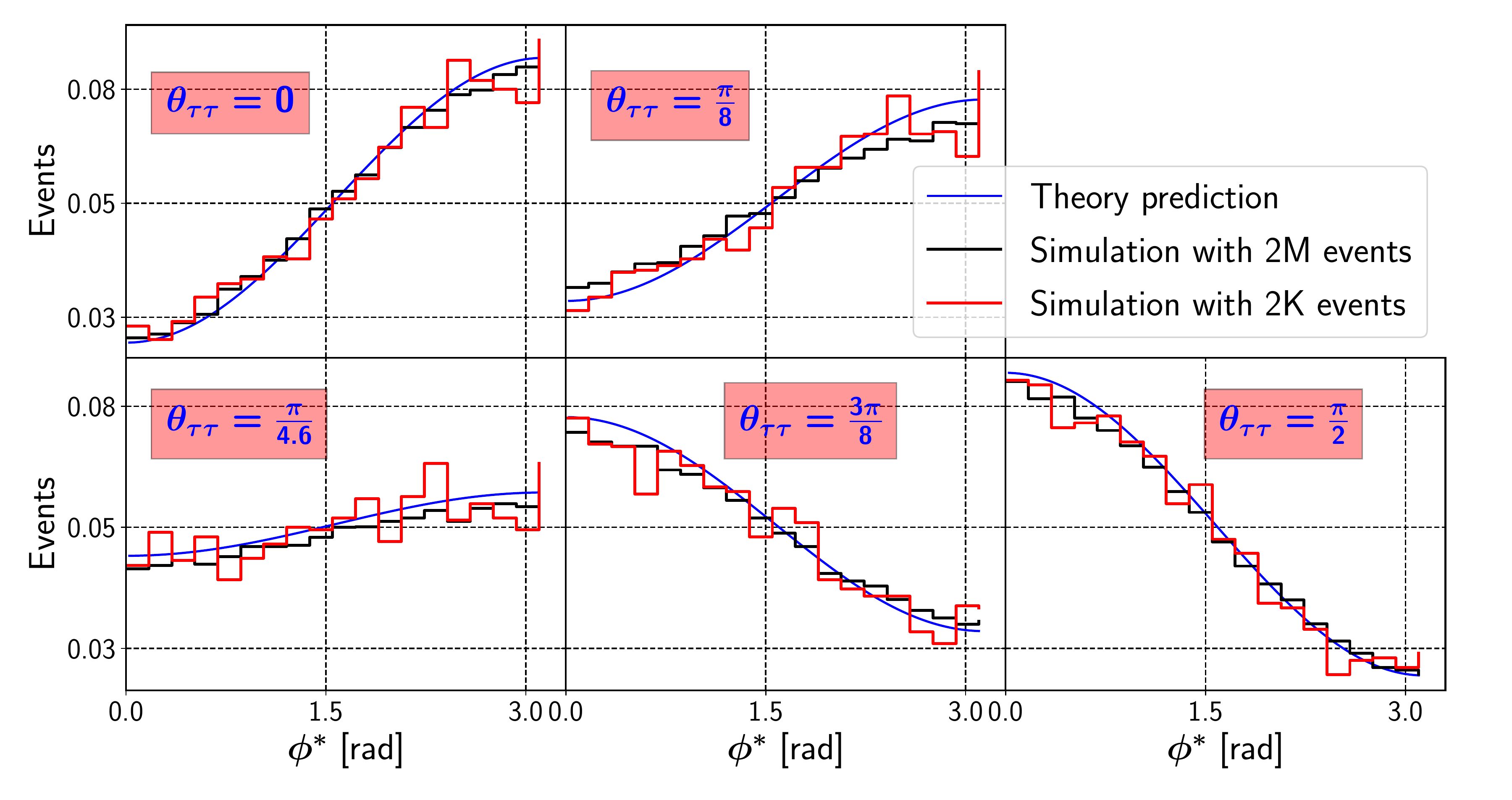}
\caption{Distributions for the events $pp \to H_2 \to \tau\bar\tau$ in the $\tau$-acoplanarity angle $\phi^*$, in the zero momentum frame, see sec.~\ref{subsec:CPviolation} for details. The red lines denote the results from a MonteCarlo simulation with MadGraph5 for the 2043 events, as expected for the chosen benchmark point at the HL-LHC. The black lines are evaluated from samples with 2M events  and indicate the infinite statistics limit. Systematic uncertainties stem from hadronisation, detector simulation, and reconstruction. The blue lines were derived from the theory prediction in eq.~\eqref{eq:theorydistribution}. For all distributions the total number of events is normalised to one. }
\label{fig:acoplanarity}
\end{figure}

We focus here on the $H_2$ boson, which in general is more strongly coupled to the SM fermions and thus yields a stronger signal, i.e.\ more events.
To separate the signal from the backgrounds, we train a Boosted Decision Tree (BDT),\footnote{We use the Tool for Multi-Variate Analysis package (TMVA) \cite{Hocker:2007ht}.} which we feed with the simulated distributions from the process $pp \to H_2 \to \tau\bar\tau$, neglecting the small contributions from $H_1$ and $H_3$.
As variables we include the invariant mass of the two reconstructed taus, the missing transverse energy and $\Delta R(\tau_{\mathrm{had}},\tau_{\mathrm{had}})$.

The BDT algorithm ranks the input variables according to their ability to separate between signal events and background events. 
The BDT classifier ranges from $-1$ to $1$ and quantifies the separability of signal and background.
Events with discriminant value near $1$ are classified as signal-like events and those near $-1$ are considered as background-like events.
The BDT response to signal and background events is shown in the left panel of fig.~\ref{fig:8} in blue and red, respectively.
The optimization of the signal significance as a function of signal and background cut efficiency is shown in the right panel of fig.~\ref{fig:8}.  The maximum cut efficiency is at BDT classifier $\ge 0.193$, corresponding to a signal significance $7\sigma$ with signal efficiency $0.57$ and background rejection efficiency $0.0059$. 
For the benchmark point with $\theta_{\tau\tau} = 0.68$, the BDT yields 2043 signal events versus 82212 background events.

Additionally, we simulate distributions for the same benchmark point but with different CP-mixing angles $\theta_{\tau\tau} = 0,\, \pi/8,\, 3/8 \pi,\, \pi/2$.
We remark that we are using the simulations with different $\theta_{\tau\tau}$ values only for comparison, and we do not check that all experimental constraints are satisfied for suitable corresponding parameter points.

As signal we consider the decay $H_2 \to \tau\bar\tau$ with subsequent decay of $\tau^\pm \to \nu_\tau \pi^\pm$.
As described above we study the tracks inside the tau jets, which carry information about the spin correlation between the tau lepton and $\pi^\pm$, and thus allow us to reconstruct the angle between the decay planes of the two $\tau$ leptons, the acoplanarity angle $\phi^\ast$ as defined above.
A $P_T$  cut on the tagged jets is applied, forcing the transverse momentum to be larger than $ 20$ GeV. 
Furthermore, we improve the quality of the events with a cut on the track impact parameter: $d_0\ge 50$ $\mu m$. 
(This cut is taken into account during the analysis and the reported numbers after BDT cut assume this cut.)
The fourvectors of the pion candidates' track are boosted to the ZMF as described in sec.~\ref{sec:discoveringCPviatautau} above.
In the ZMF the new acoplanarity angle $\phi^*$ is evaluated according to eq.~\eqref{eq:acoplanarityZMF}.

Now we turn to analysing the shape of the distribution, aiming to infer the CP-mixing angle $\theta_{\tau\tau}$ from the simulated data.
First we observe that the simulated distributions after all cuts have a very similar shape to the theory prediction for $\Gamma(\Phi)$ from eq.~\ref{eq:theorydistribution}.
We thus define the reconstructed distribution in the ZMF frame for our numerical fit to the data, introducing the fit parameters $a,\,b$, as:
\begin{equation}
\frac{1}{\Gamma} \frac{d\Gamma}{d\phi^*}({\theta_{\tau\tau}})  = a({\theta_{\tau\tau}}) - b({\theta_{\tau\tau}}) \cos\phi^*\,.
\label{eq:theory*}
\end{equation}

We find excellent agreement between our fitted values for $a_{\theta_{\tau\tau}},b_{\theta_{\tau\tau}}$ and the theoretical values in eq.~\ref{eq:theorydistribution}, which are $a_{\theta_{\tau\tau}} = 1/(2\pi)$ and $b_{\theta_{\tau\tau}} = \pi/32\left(\cos^2\theta_{\tau\tau}-\sin^2\theta_{\tau\tau}\right)$. We therefore directly compare the reconstructed distributions with the theory predictions from eq.~\ref{eq:theory_prediction}.

For our shape analysis we consider the distributions for the samples of 2043 events labelled ``2K'', corresponding to the expected event yield of the benchmark point at the HL-LHC, and the ``infinite statistics" limit labelled ``2M", corresponding to $2$ million events. 
The latter have a much smaller statistical uncertainty compared to the systematic one, which stems from uncertainties related to hadronisation, detector simulation, and the reconstruction of tau leptons.
We show the distributions for both, the small and large versions of the five signal samples, in fig.~\ref{fig:acoplanarity}.
In the figure we also show the theory prediction for $1/\Gamma d\Gamma/d\phi^*$ in eq.~\eqref{eq:theory_prediction}.

The distributions are given for $N_{\mathrm{bins}} = 20$ bins from which we create a $\chi^2$ fit for different values of $\theta_{\tau\tau}$ using 
\begin{equation}
\chi^2(\theta_{\mathrm{fit}})= \frac{\left(S_i^{\theta_{\tau\tau}} - \frac{n_S}{\Gamma} \frac{d\Gamma}{d\phi^*_i}(\theta_{\mathrm{fit}})\right)^2}{(\delta S_i)^2 + \delta_{\mathrm{syst}}^2}\,,
\label{eq:chi2}
\end{equation}
where $\theta_{\tau\tau}$ is the  mixing angle of a given benchmark point, $\theta_{\mathrm{fit}}$ an input of the theoretical distribution, $S_i^{\theta_{\tau\tau}}$ the signal distribution in bin $i$, $n_S=2043$ is the total number of signal events, and
\begin{align}
\delta S_i  = \sqrt{S_i},\  \qquad \delta_{\mathrm{syst}}  = \alpha \frac{N_{\mathrm{bkg}}}{N_{\mathrm{bins}}}\,.
\end{align}
The number of background events after the BDT cut is  $N_{\mathrm{bkg}} = 82212$ and $\alpha$ is the precision with which the background can be controlled experimentally. The background is completely flat with respect to the signal, which is an outcome of our simulation.
In the following we consider the three exemplary values $\alpha = 5\%,1\%,$ and $0.5\%$, which we assume to be conservative, realistic, and optimistic, respectively.

\begin{figure}
\centering
\includegraphics[width=0.8\textwidth]{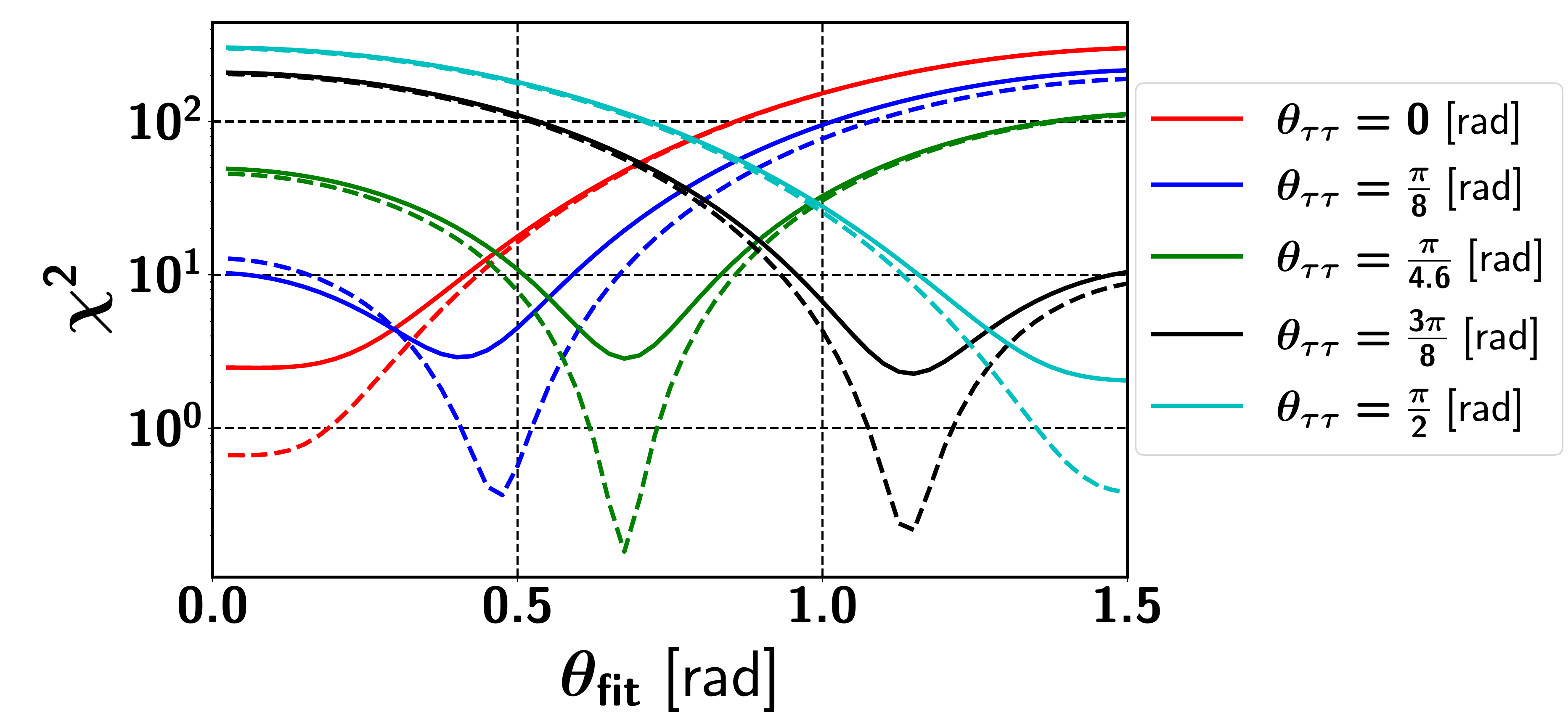}
\caption{Absolute value of the $\chi^2$ for the five different values of CP-mixing $\theta_{\tau\tau}$ evaluated according to eq.~\eqref{eq:chi2} with $\delta_{sys}=0.5\%\cdot\left(N_{bkg}/N_{bins}\right)$. The solid and dashed lines correspond to the 2K (HL-LHC) and 2M ("infinite statistics") event samples, respectively, for details see text.}
\label{fig:chi}
\end{figure}

For both, the distributions from the small and large samples, the above $\chi^2$ fit yields a minimum for $\theta_{\mathrm{fit}}$ that agrees with the set value $\theta_{\tau\tau}$ with high accuracy.
We chose the confidence level (CL) for excluding pure CP-even or CP-odd hypotheses from the $\Delta \chi^2$ distributions at 90\%.
For our 20 observables (the bins) minus the one  parameter ($\theta_{\mathrm{fit}}$) this corresponds to $\Delta\chi^2 = 27.2$.
We find that for $\alpha = 5\%$ and 1\% no statistically meaningful statement on CP violation is possible at the 90\% CL for our benchmark point at the HL-LHC.
We show the resulting $\chi^2$ distributions for the five considered CP mixing angles in fig.~\ref{fig:chi} for $\alpha = 0.5\%$.
For our benchmark point where the set value is $\theta_{\tau\tau}=\frac{\pi}{4.6}$, and considering the HL-LHC sample with 2043 events, our procedure allows to determine $\theta_{\tau\tau} \simeq \frac{\pi}{4.6} \pm 0.3$ at 90\% CL.
CP-conservation can therefore be excluded at $\gtrsim90$\% CL for this point.

\section{Conclusions}
The violation of  CP symmetry is fundamental to the baryon asymmetry of the Universe.
One of the few ways to introduce it is a CP-violating scalar sector, which implies the existence of additional scalar degrees of freedom (i.e.\ extra Higgs states) with possible observable consequences at the LHC and future colliders.
Some of the signatures that indicate the violation of CP in the scalar sector include: simultaneous observation of specific processes, top-quark associated production modes and angular momentum correlations in sequential decays such as $H_i\to ZZ\to 4\mu$ and $H_i\to\tau\bar\tau$.

In this article we consider the type I and II Two Higgs Doublet Models (THDMs) as examples for observable CP violation in the scalar sector.
We evaluated the mass eigenbasis numerically, i.e.\ without assumptions on any of the parameters.
We determined a viable parameter space region via a numerical scan over the  parameters that are compatible with the present constraints, including theoretical considerations, $B$-physics measurements, Higgs data, and measurements of electric dipole moments.
Our scan shows that the constraints allow for scalar bosons with masses of order a few hundreds  of GeV, which can be within reach of  the HL-LHC.
Moreover, we find that the possible amount of CP violation is much more suppressed in the type II THDM.

In case of CP violation the decay chain $H_i \to ZZ \to 4 \mu$ can give rise to three clearly distinct Higgs peaks. 
This can provide a clear signal for CP violation in the considered THDM (cf.\ Appendix B), where exactly two of the Higgs fields can decay to $ZZ$ at the tree-level in case of CP conservation.
However, this signature is not unambiguous, since the third resonance could stem from additional scalar fields outside the THDM.
Using the angular distributions in this decay chain turned out not to be feasible due to the coupling of the CP-odd component to $ZZ$, occurring only at loop-level, being too strongly suppressed (cf.\ Appendix A).

Towards finding an unambiguous signal of CP violation in the scalar sector we have analysed the process $pp \to H_2 \to \tau\bar\tau$  in the type I THDM at the detector level for a selected benchmark point, using a Boosted Decision Tree (BDT). 
We included the following SM backgrounds: $Z \to \tau\bar\tau$, single top and $t\bar t$, and light jet misidentification.
For our analysis the decays $\tau \to \nu\tau \pi $ were implemented. 
The detectability of CP non-conservation was quantified via a $\chi^2$ fit of the theoretically predicted distributions of the reconstructed tau-decay planes to the simulated data.
We find that CP conservation in the scalar sector can be excluded at the 90\% CL for our selected benchmark point, i.e.\ when the CP-mixing angle is close to its maximal value ($\pi/4$) and the background can be controlled with a relative accuracy of 0.5\%, which could be the accuracy target for future measurements.
Our results are conservative, since also other $\tau$-decays (such as $\tau \to \nu_\tau \rho$) can be used to study CP violation.
\subsection*{Acknowledgements}
This work has been supported by the Swiss National Science Foundation under the project number 200020/175502.
O.F.\ received funding from the European Unions Horizon 2020 research and innovation program under the Marie Sklodowska-Curie grant agreement No 674896 (Elusives). C.S. was supported by the Cluster of Excellence Precision Physics, Fundamental Interactions, and Structure of Matter (PRISMA+ EXC 2118/1) funded by the German Research Foundation (DFG) within the German Excellence Strategy (Project ID 39083149), and by grant 05H18UMCA1 of the German Federal Ministry for Education and Research (BMBF). A.H. would like to thank Waleed Esmail for fruitful discussions.

\appendix

\section{Angular correlations in $H_i \to ZZ \to 4\mu$}
\label{appendix_A}
In this appendix we investigate the process $H_i \to ZZ \to 4\mu$, and the possibility to infer the CP property of $H_i$ from angular correlations in the final state muons.
This possibility has been discussed previously, cf.\ refs.~\cite{Choi:2002jk,Buszello:2002uu,Godbole:2007cn,Chang:1993jy,Ferreira:2016jea,DeRujula:2010ys,Bolognesi:2012mm,Artoisenet:2013puc}.
Searches for such processes have been carried out by the ATLAS \cite{Aad:2015mxa} and CMS \cite{Khachatryan:2016tnr} collaborations.

\begin{figure}[!h]
\centering
\includegraphics[scale=0.5]{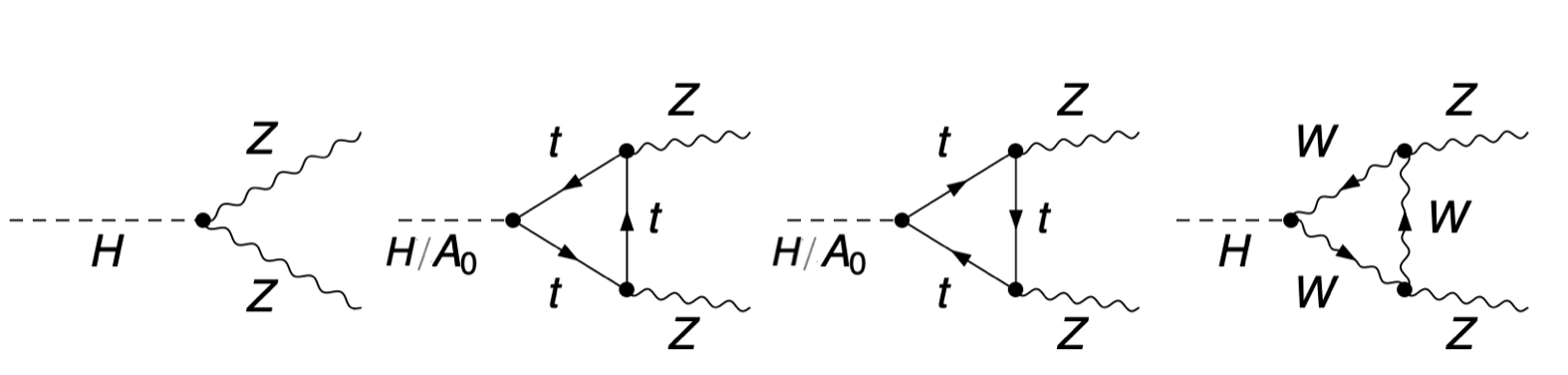}
\caption{Feynman diagrams for the coupling of CP-even ($H$) and CP-odd ($A_0$) Higgs fields to two $Z$ bosons, at tree and one-loop level.}
\label{fig:feyn}
\end{figure}

The starting point is the observation that CP-odd fields couple to $ZZ$ only at loop-level, dominantly via a loop involving a top quark, cf.\ fig.\ \ref{fig:feyn}. 
The pseudoscalar coupling to the top quark gives rise to specific correlations in the four-fermion final states.
In order to determine whether or not these final state correlations can be observed, we investigate the branching ratios of CP-even ($H$) and CP-odd scalars ($A_0$) into $ZZ$.

Let us evaluate the size of the effective couplings for $H$ and $A_0$ from the contributions in fig.~\ref{fig:feyn}.
The matrix element for Higgs decays to $ZZ$ is given by
\begin{align}
i{\mathcal{M}} &\:= \:i{\mathcal{M}^{\text{tree}}_{(H\to ZZ)}}+i{\mathcal{M}^{\text{one-loop}}_{(H/A_0\to ZZ)}} \\
 &\:= \:C_1\, \epsilon^\ast_1\epsilon^\ast_2 + C_2\, (P_2 \epsilon^\ast_1)(P_1 \epsilon^\ast_2) + C_3\, {\mathcal{E}_{\mu\nu\alpha\beta}} P^\mu_1P^\nu_2 \epsilon^{\ast\alpha}_1\epsilon^{\ast\beta}_2 \:,
\label{eq:effectivecouplings}
\end{align}
where $ \epsilon^\ast_1, \epsilon^\ast_2$ and $P_1, P_2$ are the polarization vectors and the momenta for the outgoing gauge bosons, and ${\mathcal{E}_{\mu\nu\alpha\beta}}$ is the totally antisymmetric tensor. The form factors $C_2$ and $C_3$ measure the strength of the coupling of the CP-even and CP-odd states to $ZZ$ that arises at one-loop level, while $C_1$ is the coupling of the CP-even field to $ZZ$ from the tree-level diagram.
It is the contraction of the momenta via the antisymmetric tensor $ {\mathcal{E}_{\mu\nu\alpha\beta}}$ in the last term of eq.~\eqref{eq:effectivecouplings} that gives rise to the different correlations in the four-muon final states of the process $A_0\to ZZ \to 4\mu$.
We evaluated the coefficients $C_i$ using FeynCalc\cite{Shtabovenko:2020gxv} and Package-X \cite{Patel:2015tea}. 
The tree-level and the one-loop couplings are given by:
\begin{align*}
C_1 &= \frac{i g M_Z\ \sin(\beta-\alpha)}{\cos\theta_W}\,, \\
C_2 &= \frac{i\ \sin\alpha\ g^3 m^2_t}{18 \pi^2 m_{H}^4 \sin\beta \cos\theta_W^2 m_W} \left(3 m_{H}^2+m_t^2 \ln\left(\chi_t\right)^2+\sqrt{m_{H}^4-4 m_{H}^2 m_t^2} \ln \left(\chi_t\right)\right)+\\
&\frac{i\ \sin(\beta-\alpha)\ g^3 m_W \cos\theta_W^2}{4 \pi^2 m_{H}^4 } \left(-30 m_{H}^2+(m_H^2-m_W^2) \ln\left(\chi_W\right)^2-10\sqrt{m_{H}^4-4 m_{H}^2 m_W^2} \ln \left(\chi_W\right)\right)\,,\\
C_3 &= \frac{-i g^3 m^2_t}{18 \pi^2 m_{A_0}^4\tan\beta \cos\theta_W^2 m_W} \times \\ &\left(2 m_{A_0}^2+(4\sin\theta_W^4-3\sin\theta_W^2)m^2_{A_0}\ln\left(\chi_t\right)^2-9\sqrt{m_{A_0}^4-4 m_{A_0}^2 m_t^2} \ln \left(\chi_t\right)\right)\,,
\end{align*}
where $\alpha$ is the mixing angle between the CP-even Higgs bosons, $\theta_W$ is the weak mixing angle and  
$$\chi_a =\frac{\sqrt{m_{\phi}^4 - 4m_{\phi}^2 m_{a}^2}+2m_{a}^2 - m_{\phi}^2}{2m_{a}^2}\,, $$ 
with $m_\phi$ and $m_a$ denoting the masses of the decaying Higgs bosons ($H$ or $A_0$) and of the loop particles, respectively. 

It is in principle possible to test the CP transformation property of the extra Higgs state via an asymmetry in the angular distributions of the four fermion final states \cite{Godbole:2007cn}. 
For CP-odd fields, this requires the measurement of the final state correlations in the final states from the process $A_0 \to ZZ \to 4 \mu$.
To infer the CP transformation property sucessfully, a large sample of $4\mu$ from this decay chain is needed, which in turn requires a substantial branching fraction of the process. The dominant pseudoscalar decay modes are e.g.\ $A_0\to \bar{t}t \propto\left( y_t\cos\beta\right)^2$ and $A_0\to \bar{b}b \propto \left( y_b\sin\beta\right)^2$ and also $A_0\to H^\pm W^\mp  \propto ( g_2\sin2\beta (P_A-P_W))^2$, $A_0\to H Z  \propto (\sin(\alpha-\beta)\sqrt{g_1^2+g^2_2} (P_A-P_Z))^2$.

Since the dominant decay channels are unsuppressed tree-level decays, it turns out that the branching ratio for $A_0 \to ZZ$ in THDMs is quite small, maximally about $10^{-3}$ (cf.\ \cite{Bernreuther:2009ts}), and the branching ratio to $4 \mu$ leads to a further suppression by Br$(ZZ \to 4\mu) \simeq 10^{-3}$ 
We find that the production cross section for $A_0$ is at most $0.1$ pb, which yields a total cross section for the process $pp \to A_0 \to ZZ \to 4\mu$ of $\sim 10^{-7}$ pb, and suppressing backgrounds by introducing cuts will reduce the resulting number of events that can be used for an analysis even further.
With the total luminosity at the HL-LHC being 3 (ab)$^{-1}$ it is clear that the loop-suppressed decay $A_0 \to ZZ \to 4 \mu$ is too much suppressed to use it for studying the angular correlations of the four muons. Of course the same conclusion also applies to the CP-odd component of an $H_i$ that is an admixture of a CP-even and a CP-odd field.

\section{The Higgs spectrum from $H_i \to ZZ \to 4\mu$}
\label{appendix_B}
The process $pp \to H_i \to ZZ \to 4\mu$ is a very clear channel that may contribute substantially to the discovery of the scalar $H_i$.\footnote{Another very relevant discovery channel for a scalar boson in the here considered mass range is $H_i \to 2 H_1$ \cite{Hammad:2016trm}.}
As we discussed in the previous section, it is not feasible to use the angular distributions of the final state muons from this process at the HL-LHC for establishing the existence of CP violation in the scalar sector.
However, in the context of THDMs, it can still be used to establish a signal of CP violation via the reconstructed Higgs spectrum from the invariant mass distribution of the Higgs decay products.

\begin{figure}[!h]
\centering
\includegraphics[scale=0.5]{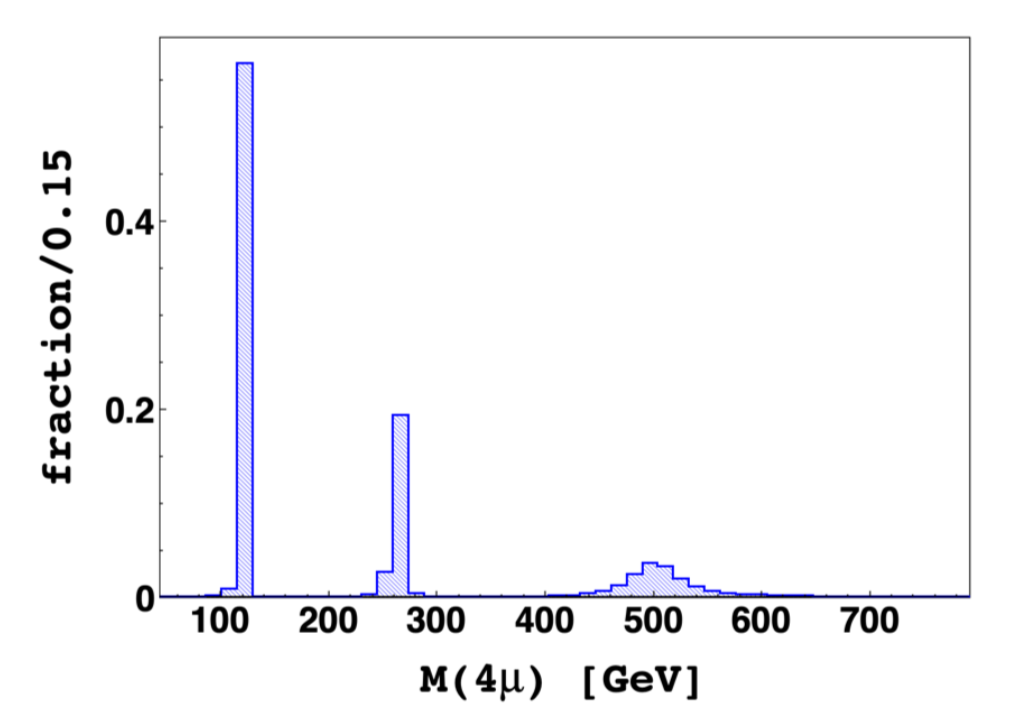}
\caption{Distribution of the total invariant mass of the four muon final state from the process $pp\to H_i\to ZZ\to 4\mu$, from an inclusive simulation of the signal sample with 20M events, including a fast detector simulation with Delphes.}
\label{fig:A7}
\end{figure}

When the scalars are not pure eigenstates of CP, all of the $H_i$ can have sizeable branching ratios into $ZZ$, giving rise to three resonances in the $4\mu$ final state, as shown in fig.~\ref{fig:A7}.
Measuring three peaks for the invariant masses of the four muon final states is thus a clear signal for CP violation within the complex THDM.

\begin{figure}[h!]
	\centering
	\includegraphics[scale=0.25]{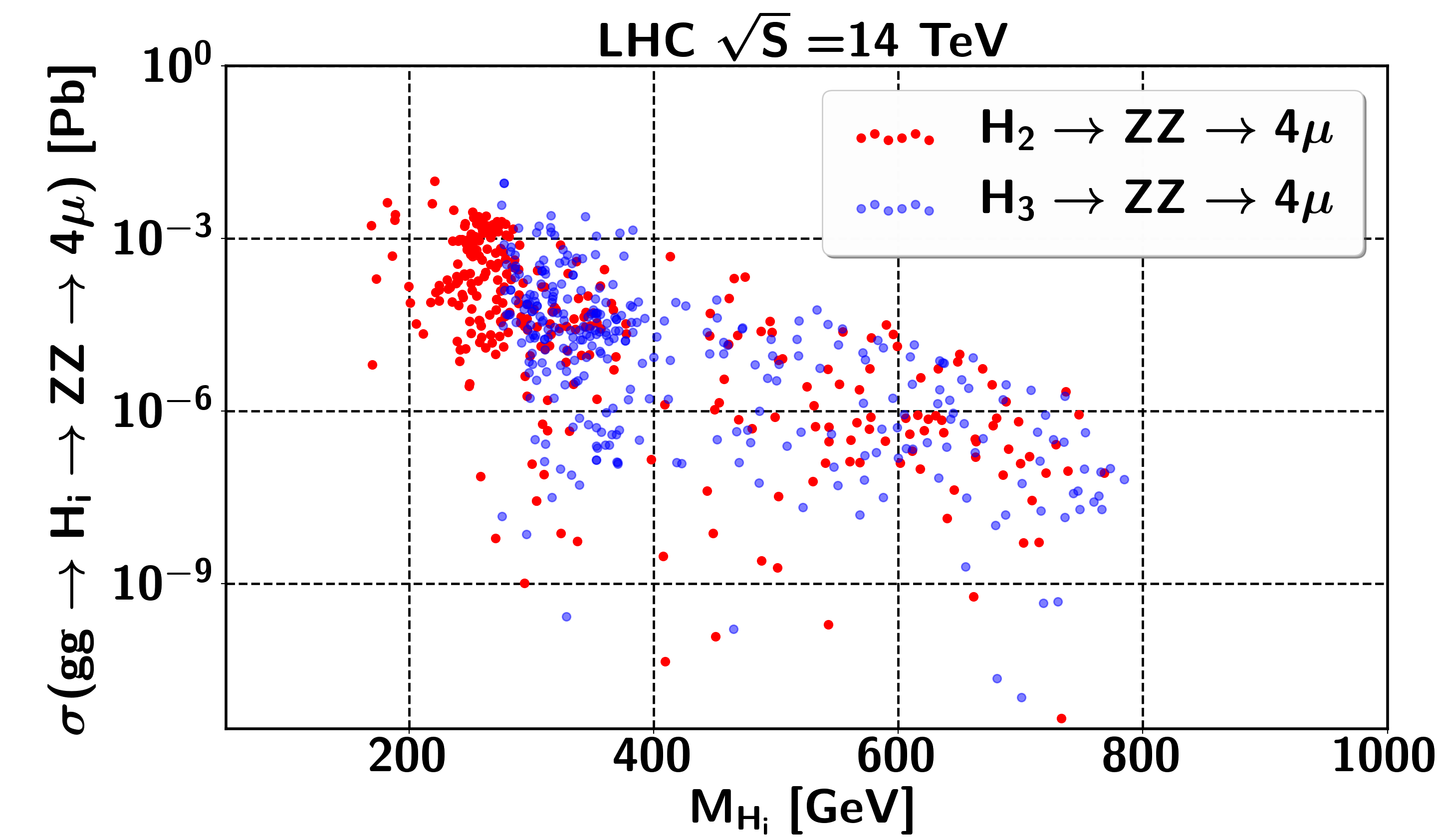}
	\caption{Total cross sections for the process $pp\to H_i\to ZZ \to 4\mu$ at the LHC with $\sqrt{s}=14$ TeV. The scatter plot uses the results from the parameter space scan in sec.\ \ref{sec:scan}. 
	}
	\label{fig:HtoZZ-xsec}
\end{figure}
To evaluate the observability of this process, we consider a benchmark point with $m_{H_2}=260$ GeV and $m_{H_3}=500$ GeV, based on the model parameters $\tan\beta = 4$,  $\lambda_1= 0.172,\ \lambda_2=0.0828,\ \lambda_3=5.149,\ \lambda_4=-0.313,\  \Re(\lambda_5)= -4.6431, \eta(\lambda_5)=0.81$ and $m_{12}^2=1.091\times10^{4}$ GeV$^2$.
The parameters $m^2_{11}$ and $m^2_{22}$ are then fixed by the previous parameters due to the tadpole equations, cf.\ eq.~\eqref{eq:tadpole-equations}.
In fig.~\ref{fig:HtoZZ-xsec} we show the total cross section for the process $pp\to H_i\to ZZ\to 4\mu$ for number of scanned points from our scan in sec.\ \ref{sec:scan}.

We consider the following backgrounds:
The dominant SM background that contributes to the final state with $4\mu$ is $ZZ$ production. 
Other reducible backgrounds are $WW$ and $WZ$ production, where one of the jets is misidentified as a muon. This set of background processes can be sufficiently reduced by the requirement of tight isolation criteria for the hard final state muons. The set of backgrounds with $tt$ production and the associated production of top quark with a $W$ boson can be reduced by vetoing $b$-jets. The last set of backgrounds with three gauge boson production is highly suppressed by the large missing energy associated to these processes and will not be included in the analysis. All the considered and included backgrounds are listed with their cross sections in tab.~\ref{tab:backgroundsZZ}.

\begin{table}
\begin{center}
\begin{tabular}{|l|c|}
\hline
Backgrounds & $\sigma_{(HL\text{-}LHC)} [Pb]$   \\ \hline\hline
$pp\to ZZ\to 4\mu$  & 0.0065  \\ \hline
$pp\to \bar{t}t, \quad \mbox{where}  \; t\to \text{leptons}$  &6.7  \\ \hline
$pp\to \bar{t}t Z$  & 0.0002   \\ \hline
$pp\to WZ\to 3\mu+\nu_\mu$  & 0.099   \\ \hline
$pp\to t W b, \quad \mbox{where}  \; t\to \text{leptons}$  &7.1   \\ \hline
\end{tabular}
\end{center}
\caption{Dominant background processes considered in our analysis and their total cross sections. The samples have been produced with the following cuts: $P_T(j)\ge 20$ GeV, $P_T(l)\ge 10$ GeV. }
\label{tab:backgroundsZZ}
\end{table}

\begin{figure}
\centering
\includegraphics[width=7cm,height=4.8cm]{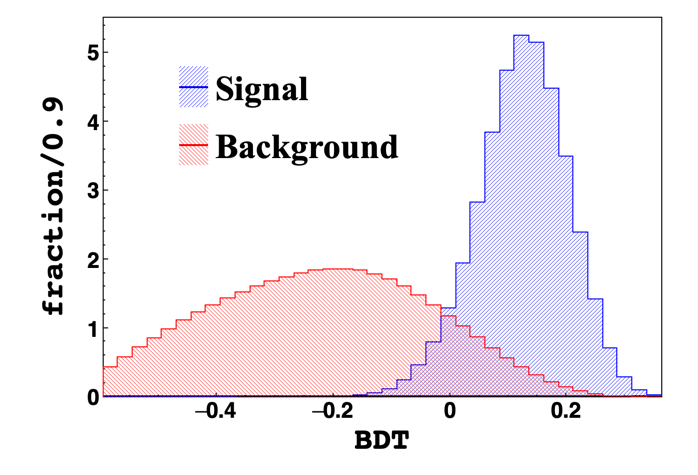}~~\includegraphics[width=8cm,height=5cm]{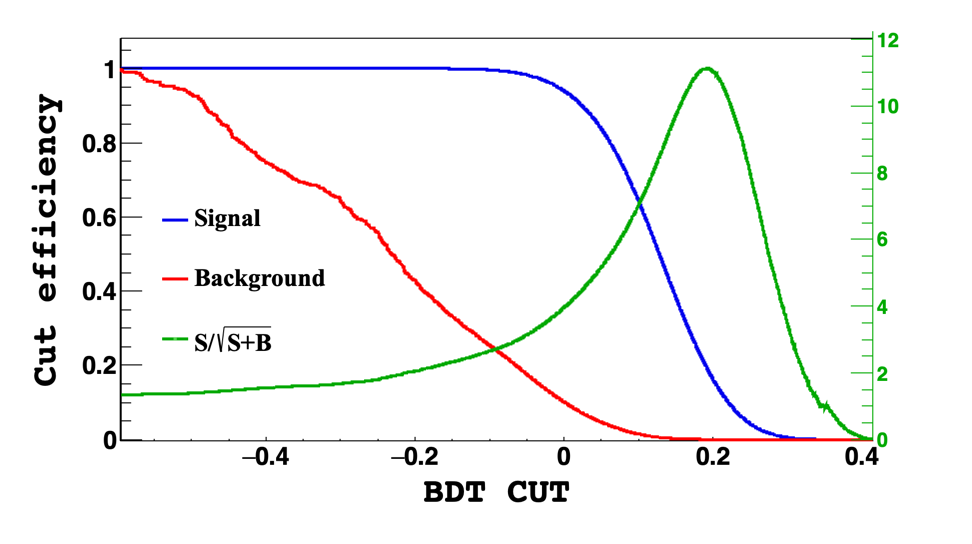}

\caption{{\it Left:} The distribution of the Boosted Decision Tree response to the signal (blue) and to the background (red), superimposed.
{\it Right: } Cut efficiency as a function of the BDT cut. For a cut value greater than $0.193$ one can get $S/\sqrt{S+B}=11\sigma$ with number of signal events $= 939$ and background events $= 6185$. The cut efficiency for the signal is $0.187$ and for the background $0.0004$.}
\label{fig:A8}
\end{figure}

We constructed all possible kinematic variables  for the signal and all relevant backgrounds and used the BDT to optimize the signal to background classifier as shown in fig.~\ref{fig:A8} (left). According to the BDT ranking, the invariant mass of the four final state muons is the most important variable to separate the signal from the backgrounds. 
The fact that all three neutral bosons can decay into a pair of $Z$ bosons proofs that our benchmark point has Higgs states with mixed CP properties, since otherwise one of the three bosons would be a pure pseudoscalar which does not interact with $ZZ$ at tree-level. 

The signal significance as a function of signal and background cut efficiency is shown in the right panel of fig.~\ref{fig:A8}. The maximum cut efficiency is at BDT $\ge 0.193$, corresponding to a signal significance 11$\sigma$ with signal efficiency 0.187 and background rejection efficiency 0.0004, which demonstrates an excellent discovery potential for our benchmark point in this channel alone.

We emphasize that the observation of three scalar resonances in the $4\mu$ final state is a positive signal for CP violation {\it only} in the THDM, because there it is absent when CP is conserved. 
It is not an unambiguous signal of CP violation outside the THDM, since the third resonance could stem from some other CP-even scalar field.


\end{document}